\documentclass[aps,longbibliography,showpacs,twocolumn,superscriptaddress,amsmath,amssymb,verbatim]{revtex4-2}
\usepackage[thinlines]{easytable}
\usepackage{graphicx}
\usepackage{lmodern}
\usepackage{epstopdf}
\usepackage{dcolumn}
\usepackage{bm}
\usepackage{subfigure}
\usepackage{makecell}
\usepackage{amsmath} 
\usepackage[pdftex,colorlinks=true,citecolor=blue,linkcolor=blue,urlcolor=blue,bookmarks=true]{hyperref}

\usepackage[utf8]{inputenc}

\usepackage{xcolor}
\usepackage{tikz}

\usepackage{color}
\usepackage{textcomp}
\definecolor{red}{rgb}{1,0,0}

\definecolor{blue}{rgb}{0,0,1}

\definecolor{green}{rgb}{0,1,0}

\begin{document}
	\preprint{APS}

\title{\textcolor{black}{Coexistence of anomalous spin dynamics and weak magnetic order  in a chiral trillium lattice K$_{2}$FeSn(PO$_{4}$)$_{3}$}}
\author{J. Khatua}
\thanks{These authors contributed equally to this work.}
\affiliation{Department of Physics, Sungkyunkwan University, Suwon 16419, Republic of Korea}
\author{S. Krishnamoorthi}
\thanks{These authors contributed equally to this work.}
\affiliation{Institute of Physics, Academia Sinica, Taipei 11529, Taiwan}
\author{Changhyun Koo}
\affiliation{Department of Physics, Sungkyunkwan University, Suwon 16419, Republic of Korea}
\author{Gyungbin Ban}
\affiliation{Department of Physics, Sungkyunkwan University, Suwon 16419, Republic of Korea}
\author{Taeyun Kim}
\affiliation{Department of Physics, Sungkyunkwan University, Suwon 16419, Republic of Korea}
\author{Suyoung Kim}
\affiliation{Department of Physics, Simon Fraser University, Burnaby, BC V5A 1S6, Canada}
\author{Yugo Oshima}
\affiliation{RIKEN Pioneering Research Institute, Wako, Saitama 351-0198, Japan}
\author{Jonas A. Krieger}
\affiliation{
PSI Center for Neutron and Muon Sciences CNM, 5232 Villigen, Switzerland}
\author{Thomas J. Hicken}
\affiliation{
	PSI Center for Neutron and Muon Sciences CNM, 5232 Villigen, Switzerland}
\author{Hubertus Luetkens}
\affiliation{
	PSI Center for Neutron and Muon Sciences CNM, 5232 Villigen, Switzerland}
\author{Marc Uhlarz}
\affiliation{Dresden High Magnetic Field Laboratory (HLD-EMFL), Helmholtz-Zentrum Dresden-Rossendorf, 01328 Dresden, Germany}
\author{Eundeok Mun}
\affiliation{Department of Physics, Simon Fraser University, Burnaby, BC V5A 1S6, Canada}
\author{Kyeong Jun Lee}
\affiliation{Department of Physics, Chung-Ang University, Seoul 60974, Republic of Korea}
\author{Seo Hyoung Chang}
\affiliation{Department of Physics, Chung-Ang University, Seoul 60974, Republic of Korea}
\author{R. Sankar}
\email[]{sankarndf@gmail.com}
\affiliation{Institute of Physics, Academia Sinica, Taipei 11529, Taiwan}
\author{Kwang-Yong Choi}
\email[]{choisky99@skku.edu}
\affiliation{ Department of Physics, Sungkyunkwan University, Suwon 16419, Republic of Korea}

\date{\today}

\begin{abstract}
 Trillium lattices, where magnetic ions form a three-dimensional chiral network of corner-sharing equilateral triangular motifs, offer a prominent platform to explore exotic quantum states. In this work, we report ground-state properties of the $S$ = 5/2 trillium lattice compound K$_{2}$FeSn(PO$_{4}$)$_{3}$ through thermodynamic, electron spin resonance (ESR), and muon spin relaxation (${\mu}$SR) experiments. Thermodynamic and ESR measurements reveal the two-step evolution of magnetic correlations across $T^{*}$ = 11 K, which results from an interplay between dominant antiferromagnetic Heisenberg interactions and subleading interactions. Below $T^{*}$, \textit{dc} and \textit{ac} magnetic susceptibilities indicate weak \textcolor{black}{magnetic ordering} at $T_{\rm N} \approx 2$ K under low fields, which is suppressed for $\mu_{0}H \geq 2$ T, consistent with a power-law dependence of magnetic specific heat at low temperatures.  $\mu$SR experiments confirm the dominance of persistent spin dynamics and the absence of conventional spin freezing, supporting the subtle nature of weak magnetic ordering coexisting with \textcolor{black}{ strong  spin fluctuations}. These findings underscore the potential for realizing a classical spin-liquid ground state with exotic excitations in high-spin trillium lattice systems.	\end{abstract}
\maketitle
\section{INTRODUCTION}
 Geometrically frustrated magnets are of great interest in condensed matter physics as they offer a viable platform for realizing various quantum many-body phenomena, including quantum spin liquids (QSLs), spin ices, and fractional  quantum Hall
 liquids \cite{Balents2010,Savary_2016,annurev38,Lau2006,bacciconi2024theoryfractionalquantumhall}. QSLs are elusive states of matter characterized by  the absence of long-range magnetic order (LRO) even as $T\rightarrow 0$ K, despite stong magnetic interactions, and host fractionalized excitations such as spinons and Majorana fermions \cite{KHATUA20231,Takagi2019,scienceaay0668,10.1093/ptep/ptad115,Sears2020}.\\ 
 While QSLs are well understood in one-dimensional (1D) systems \cite{PhysRevB.52.13368,PhysRevLett.111.137205}, their realization in higher dimensions remains challenging. Following Anderson’s proposal of a QSL on the triangular lattice \cite{ANDERSON1973153} and Kitaev’s model on the honeycomb lattice \cite{Kitaev_2006}, numerous 2D QSL candidates have been identified \cite{Jeon2024,Khuntia2020,PhysRevLett.133.266703,qute,Arh2022,Bordelon2019,Sears2020}. Interestingly, despite weaker quantum fluctuations in 3D, hyperkagome and pyrochlore lattices exhibit QSL-like signatures \cite{Chillal2020,PhysRevLett.116.107203,Plumb2019,Gao2019}. Identifying new 3D QSL materials could deepen our understanding of spin liquids and distinguish them from their 2D counterparts.  \\
Notably, different types of spin liquids emerge depending on spin-lattice symmetry \cite{PhysRevB.65.165113}. One salient possibility is a chiral spin liquid (CSL), which, despite lacking LRO, breaks time-reversal and parity symmetries \cite{PhysRevLett.59.2095,PhysRevB.106.094417}. The CSL state arises from chiral interactions like $\textbf{\textit{S}}_i \cdot (\textbf{\textit{S}}_j \times \textbf{\textit{S}}_k)$, involving spins at the vertices of a triangle \cite{PhysRevB.109.125146}. Theoretically, the Heisenberg model with chiral interactions   supports the CSL state in 2D frustrated lattices for $S$ = $1/2$ \cite{PhysRevB.110.L041113,PhysRevB.109.125146,Bauer2014,PhysRevB.103.L041108}, while for $S \to \infty$, chiral interactions in 3D lattices could engender a classical CSL with fracton excitations \cite{Lozano2024}.
The classical spin liquid state is characterized by an extensive ground-state degeneracy and can exhibit either algebraic or exponential spin correlations \cite{PhysRevB.111.134413}. In 3D rare-earth pyrochlores, this brings about the emergence of a Coulomb phase with exotic monopole excitations and a U(1) gauge field \cite{PhysRevB.110.L020402,Niggemann_2020,PhysRevLett.93.167204,annurev,annurev38}. However, discovering suitable candidates to explore the classical CSL state is highly sought-after. \\In this vein, newly discovered 3D trillium lattices provide a promising platform for delving into chiral phenomena, driven by their non-centrosymmetric chiral spin topology \cite{PhysRevLett.127.157204,PhysRevB.74.224441,10.1063/5.0096942,li2024classificationspin12fermionicquantum,Gonzalez2024,PhysRevB.109.184432,PhysRevLett.131.146701,Kub2024,PhysRevB.110.224405}. Trillium materials, such as MnSi and EuPtSi, which crystallize in the non-centrosymmetric space group $P2_{1}3$, are found to host  magnetic skyrmions \cite{science1166767,Kakihana2017} and chiral phonons \cite{mahraj2024chiralphononicelectronicedge}.  Furthermore, trillium lattices, where classical spins form a 3D chiral network, have been proposed to manifest a spin ice-like classical spin liquid state, where a ``two-in and one-out" configuration is satisfied on each triangle, contrasting with the ``two-in and two-out" ice rules in pyrochlore lattices \cite{PhysRevB.82.014410,PhysRevB.78.014404,PhysRevLett.128.177201}.\\
Herein, we report the ground-state properties of single-crystal K$_{2}$FeSn(PO$_{4}$)$_{3}$ (hereafter KFSPO), a $S$ = 5/2 3D trillium lattice compound, through thermodynamic, ESR, and $\mu$SR measurements.
Magnetic susceptibility, specific heat, and ESR reveal a two-stage evolution of magnetic correlations across $T^{*} \approx$ 11 K.
Below $T^{*}$, \textit{dc} and \textit{ac} magnetic susceptibilities indicate weak magnetic ordering at $T_{\rm N} \approx 2$ K under low fields, which is suppressed for $\mu_{0}H \geq 2$ T, consistent with the weak field dependence of magnetic specific heat at low temperatures. $\mu$SR measurement confirms the dominance of  persistent spin dynamics in both zero-field and 3.4 T, suggesting \textcolor{black}{a ground state where dynamically fluctuating spins coexist with weak quasistatic local magnetic fields.}
\section{EXPERIMENTAL DETAILS}
 High-quality  single crystals of dipotassium iron(III) tin(IV) tris­(orthophosphate), K$_{2}$FeSn(PO$_{4}$)$_{3}$, were successfully grown using a self-flux method within the pseudo-quaternary system K$_{2}$O-P$_{2}$O$_{5}$-Fe$_{2}$O$_{3}$-SnO$_{2}$. The stoichiometric amounts of SnO$_{2}$ (6.0 mmol), Fe$_{2}$O$_{3}$ (9.0 mmol), KPO$_{3}$ (5.62 mmol), and K$_{4}$P$_{2}$O$_{7}$ (3.1 mmol) were thoroughly ground and placed in a platinum crucible. The reactants were heated to 1150$^\circ$C over 10 h and maintained at this temperature for 24 h. At this stage, the melt dissolved the metal oxides and was subsequently cooled to 750$^{\circ}$C at a rate of 2$^{\circ}$C per hour. After natural cooling to room temperature, pink-colored single crystals of  KFSPO, predominantly tetrahedral in shape, were obtained from the remaining flux by leaching with deionized water \cite{Zatovskywm2157}.\\ Powder
x-ray diffraction (XRD) measurements using a BrukerAXS (D8-Advance) x-ray diffractometer with Cu K$_{\alpha}$
radiation ($\lambda$ = 1.54 \AA) were performend on the crushed single crystals of
 KFSPO at room temperature. While the diffraction pattern for x-ray beam perpendicular to the (lll) planes was obtained from high-resolution XRD (HR-XRD, Bruker AXS D8) with Cu K$_{\alpha}$ beam source.  Detailed results of the XRD are presented in Supplementary Note 1 \cite{sm}.
\\ 
The \textit{dc} magnetic susceptibility and magnetization measurements were performed using a superconducting quantum interference device vibrating sample magnetometer (SQUID-VSM, Quantum Design,
USA) in the temperature range 2 K $\leq$ $T$ $\leq$ 300 K
and in magnetic fields up to 7 T. \textcolor{black}{In addition, low-temperature \textit{dc} magnetic susceptibility, magnetization, and \textit{ac} magnetic susceptibility measurements were carried
out down to 0.4 K} using the 3He option of the Quantum Design MPMS
SQUID system.\\ High-field magnetization measurement was conducted
at Dresden High Magnetic Field Laboratory, sweeping a
magnetic field up to 55 T at 1.3 K using a nondestructive
pulsed magnet along an arbitrary direction with respect to the crystal axes \textcolor{black}{with the crystal orientation not precisely determined}. The obtained data were scaled to isothermal magnetization measurements obtained with a SQUID-VSM at the same temperature. Orientation-dependent magnetic susceptibilities are presented in Supplementary Note 2 \cite{sm}. \\
Specific
heat measurements were performed using a standard relaxation method with a physical property measurement
system (PPMS, Quantum Design, USA) in the temperature range 2 K $\leq$ $T$ $\leq$ 300 K and in magnetic fields up
to 9 T. Furthermore, specific heat was measured in the
temperature range  0.4 K $\leq$ $T$ $\leq$ 4 K using 3He option at Quantum
Design Dynacool PPMS in several magnetic fields up to 9 T. \\   ESR measurements were performed using a conventional X-band ($f$ = 9.12 GHz) ESR spectrometer (JEOL, JES-RE3X) in RIKEN. A single crystal of KFSPO was loaded on a quartz rod, applying the magnetic field along the [111] crystal orientation. A continuous $^{4}$He-flow cryostat enabled controlling temperatures from 3.8 K to 280 K for the experiments.\\
 Muon spin relaxation ($\mu$SR) experiments were conducted
on the FLAME spectrometers at Paul Scherrer Institut
(PSI) in Villigen, Switzerland under zero-field (ZF)
and longitudinal field (LF) configurations. The initial muon spin direction was oriented along the beam direction.   Many small
crystals of  KFSPO  were mounted onto the
sample holder using GE varnish and enclosed between two layers of 25 $\mu$m thick copper foil.
A Variox system combined with a Kelvinox dilution fridge insert enabled measurements across a temperature range 30 mK $\leq$ $T$ $\leq$ 100 K and  a LF range of 0 T $\leq$ $\mu_{0}H$ $\leq$ 3.4 T.  The time evolution of the muon spin asymmetry at selected temperatures under an applied LF of $\mu_{0}H$ = 3.4 T is shown in  Supplementary Note 4 \cite{sm}. \textcolor{black}{All the $\mu$SR spectra were modeled with a constant background B${\rm g} \approx 0.013$, which remained nearly independent of both temperature and applied field.}
The $\mu$SR data were analyzed using the musrfit software \cite{SUTER201269}. Although high-quality single crystals of  KFSPO were used, the enantiomeric purity of individual crystals was not verified. Since multiple crystals were measured simultaneously in the $\mu$SR experiments, it is likely that both enantiomers were present.
\begin{figure*}
	\centering
	\includegraphics[ width=\textwidth]{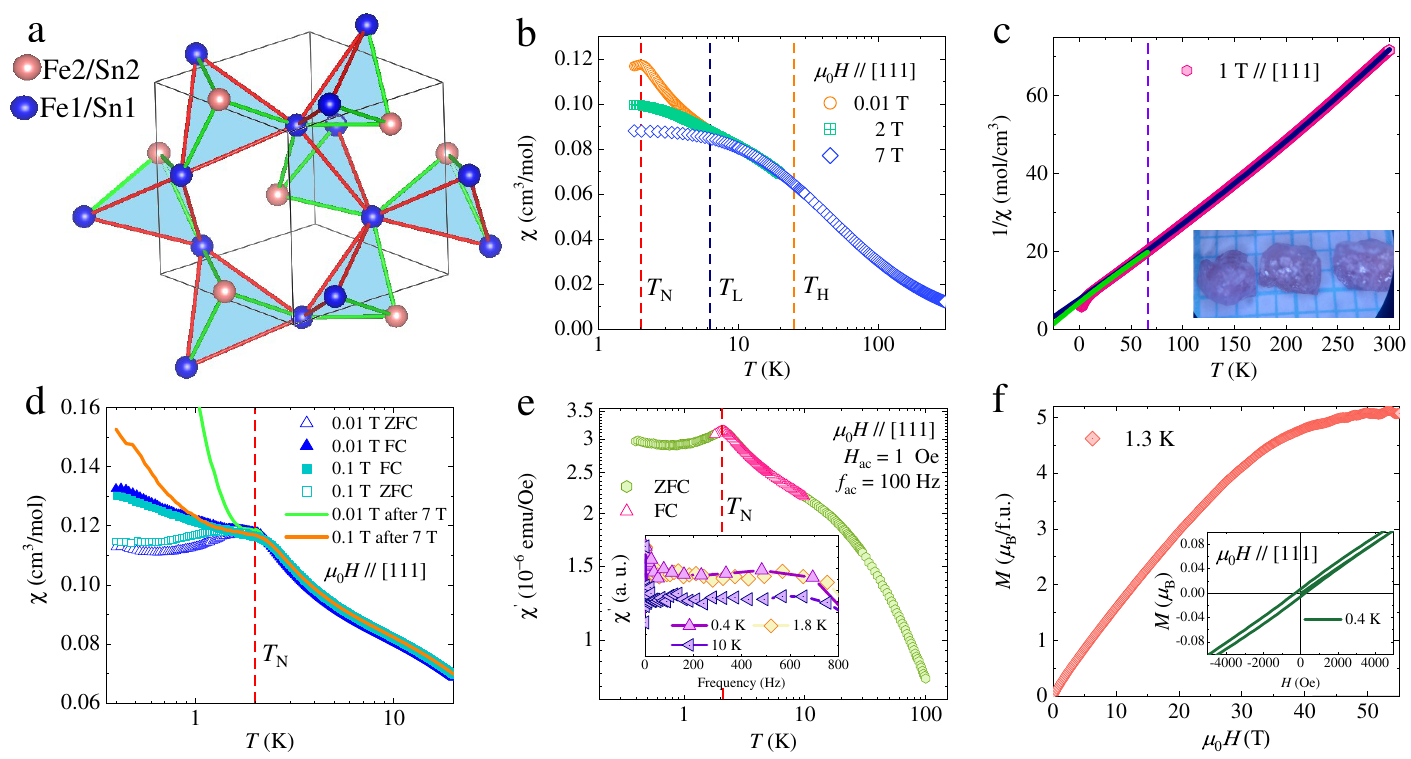}
	\caption{(a) Featuring a hypertrillium lattice formed by two magnetic sites of Fe$^{3+}$ ions, with a green bond length (Fe${2}$--Fe${1}$) of 4.96 Å and a red bond length (Fe${1}$--Fe${1}$) of 6.10 Å. (b) Temperature dependence of zero-field-cooled (ZFC) magnetic susceptibility at several magnetic fields applied parallel to the [111] direction. The dashed vertical lines around
		$T_{\rm L}$ = 6.3 K and $T_{\rm H}$ = 25 K indicate two characteristic features of magnetic correlations, and  weak magnetic ordering at $T_{\rm N}$ $\approx$ 2 K, as described in the text. (c) Temperature dependence of inverse magnetic susceptibility with a Curie-Weiss (CW) fit in two temperature regions, where deviation from the high-temperature CW fit  is marked  by a dashed vertical line around 66 K. Inset shows a photo of the single crystals. (d) Temperature dependence of ZFC and FC susceptibility at several fields in the temperature range 0.4 K $\leq$ $T$ $\leq$ 20 K. (e) Temperature dependence of real part of \textit{ac} magnetic susceptibility ($\chi^{'}$) at frequency $f_{\rm ac}$ = 100 Hz. Inset shows frequency dependence of normalized $\chi^{'}$ at three different temperatures. (f) Isotherm magnetization as a function of magnetic field up to 55 T, \textcolor{black}{with the field applied in an arbitrary orientation relative to the crystal axes}. Inset shows the small hysteresis loop in magnetization isotherm at  0.4 K. } {\label{sus}}
\end{figure*} 
\section{RESULTS}	
\subsection{Crystal structure and x-ray diffraction} To confirm the phase purity of KFSPO, x-ray diffraction (XRD) measurements were performed on crushed single crystals. Rietveld refinement of the XRD (see Supplementary Note 1 \cite{sm} and Fig.~1) data reveals that the compound crystallizes in the cubic space group $P2_{1}3$ with no detectable impurity phase \cite{Tobyhw0089}. The obtained atomic coordinates and lattice parameters are consistent with the previous report \cite{Zatovskywm2157}.   Fe$^{3+}$ ions occupy two magnetic sites, each forming a trillium lattice, which is a chiral network of three equilateral triangular lattices (see Supplementary \cite{sm} Fig.~1(b)). In the Fe$1$ trillium lattice, the exchange coupled  Fe$1$-Fe$2$ bond length of 4.96 $\text{\AA}$  creates a 3D network of corner-sharing
tetrahedra, known as the hypertrillium lattice, shown in Fig.~\ref{sus}(a). Like its sister compound K$_{2}$CrTi(PO$_{4}$)$_{3}$ \cite{PhysRevB.109.184432}, KFSPO exhibits unavoidable magnetic site dilution due to the similar atomic radii of Fe$^{3+}$ (0.64\,Å) and Sn$^{4+}$ (0.62\,Å) ions.  Indeed, our XRD analysis confirms a significant degree of site dilution, with the Fe1 site 44(3)\text{\%} occupied by Fe and 56(2)\text{\%} by Sn, and the Fe2 site 55(3)\text{\%} Fe and 45(1)\text{\%} Sn (see Supplementary Note 1 \cite{sm}). Nevertheless, the remaining magnetic Fe ions are effectively interconnected through the Fe1–Fe2 network to sustain a 3D exchange topology. Even with partial Sn substitution, the Fe network maintains the connectivity of the hypertrillium lattice, enabling magnetic properties comparable to those seen in pristine magnetic Fe-based trillium systems \cite{10.1063/5.0096942}. As recently proposed \cite{Gonzalez2024}, a key question is whether the dynamic spin correlations are intrinsically tied to the unique geometry of the hypertrillium lattice in the double trillium lattice systems. In this context, the observation of a dynamic ground state despite substantial site dilution may point to the intrinsic robustness of the hypertrillium spin topology.
\subsection{Magnetic susceptibility}
Figure~\ref{sus}(b) shows the temperature dependence of zero-field-cooled (ZFC) \textit{dc} magnetic susceptibility ($\chi(T)$) at several magnetic fields applied parallel to the [111] direction. Upon lowering the temperature, $\chi(T)$ exhibits a broad feature at $T_{\rm H}$ = 25 K (dashed orange line) with a subsequent increase below $T_{\rm L}$.   The broad feature indicates short-range spin correlations driven by thermal fluctuations in a system dominated by Heisenberg interactions \cite{PhysRevB.86.205134}.
On the other hand, the low-$T$ increase is likely due to subdominant \textcolor{black}{ferromagnetic interactions similar to that observed in the isostructural compound  K$_{2}$CrTi(PO$_{4}$)$_{3}$ \cite{PhysRevB.109.184432}}.\\ Upon
further lowering temperature below $T_{\rm L}$, $\chi{(T)}$ exhibits a weak kink at $T_{\rm N}$ $\approx$ 2 K under $\mu_{0}H$ = 0.01 T, alluding to the occurrence of magnetic ordering. Moreover, for magnetic fields 
\mbox{$\mu_{0}H \geq 2$ T}, the positions of 
$T_{\rm L}$ 
and $T_{\rm H}$ remain unchanged, while the weak kink around $T_{\rm N}$
is suppressed, triggering 
$\chi(T)$ to saturate at low temperatures (see \cite{sm}). This indicates that applying a magnetic field of $\mu_{0}H \geq 2$ T quenches  the weak ferromagnetic moments.
\\ In order to estimate dominant exchange interactions between $S$ = 5/2 spins of Fe$^{3+}$ ions in KFSPO, the inverse magnetic susceptibility (1/$\chi(T)$) data  were fitted  by the Curie-Weiss (CW) law, $\chi(T)$ = $\chi_{0}$ $+$ $C/(T-\theta_{\rm CW})$. Here, $\chi_{0}$ is the  sum of temperature-independent contributions
of diamagnetic core susceptibility and Van-Vleck susceptibility, $C$ is the Curie constant, and $\theta_{\rm CW}$ represents the CW temperature. The high-temperature CW fit (100 K $\leq$ $T$ $\leq$ 300 K, navy line in Fig.~\ref{sus}(c)) yields $\chi_{0}$ = $-2.56$ $\times 10^{-3}$  cm$^3$/mol, \mbox{$C$ = 5.66(3) cm$^{3}$ K/mol}, and \mbox{$\theta_{\rm CW}$ = $-$ 43(2) K}.  The effective magnetic moment, $\mu_{\rm eff} = \sqrt{8C}$ =  6.72(4) $\mu_{\rm B}$, is somewhat higher than the theoretical value of  $\mu_{\rm eff}^{\rm theo}$ = 5.92 $\mu_{\rm B}$ for Fe$^{3+}$ ($S$ = 5/2), indicating the presence of moderate orbital contributions, similar to those observed in other trillium lattice compounds \cite{Kub2024}.   The obtained negative CW temperature suggests the presence of dominant antiferromagnetic interactions. Importantly, below 66\,K, $\chi(T)$ deviates from the high-temperature Curie--Weiss fit and follows a secondary Curie--Weiss regime with $\theta_{\rm CW} = -30$\,K, \textcolor{black}{which is less negative than the high-temperature value ($\theta_{\rm CW} = -43$\,K). This reduction suggests that subdominant ferromagnetic interactions become active at lower temperatures alongside dominant antiferromagnetic interactions
 (see below) \cite{PhysRevB.109.184432}}. \textcolor{black}{Alternatively, this behavior could also arise from the gradual buildup of renormalized antiferromagnetic spin correlations, even in the absence of subdominant ferromagnetic exchange interactions.}\\
To understand the nature of the weak kink around $T_{\rm N}$, ZFC and FC susceptibility measurements were performed  down to 0.4 K under two magnetic fields as shown in Fig.~\ref{sus}(d). The ZFC-FC bifurcation 
at $\mu_{0}H$ = 0.01 T implies the presence  
of ferromagnetic moments below $T_{\rm N}$, while the reduced bifurcation at $\mu_{0}H$ = 0.1 T  reflects the suppression of ferromagnetic moments with increasing field. Strikingly, field cycling to 7 T at 0.4 K alters $\chi{(T)}$ at 0.01 T (labeled as ``after 7 T'' in Fig.~\ref{sus}(d)), suggesting high-field-aligned ferromagnetic components retain partial alignment after reducing the field, yielding distinct magnetic responses. The absence of frequency dependence in the \textit{ac} magnetic susceptibility (inset of Fig.~\ref{sus}(e)) above and below  $T_{\rm N}$ rules out conventional spin freezing, confirming that the irreversibility in 
$\chi(T)$ arises from a weak ferromagnetic moment \cite{PhysRevB.109.184432}. \textcolor{black}{Even though our \textit{ac} magnetic susceptibility measurements are limited to low frequencies, the absence of conventional spin freezing or spin glass is further supported by our $\mu$SR experiments (see sec.\ref{frr}) \cite{Sibille2017}.} This also indicates that magnetic site dilution in the three-dimensionally connected, double trillium lattices gives rise to novel 3D frustrated spin networks, without introducing exchange randomness or quenched disorder. Additionally, the weak kink in the real part of \textit{ac} magnetic susceptibility (Fig.~\ref{sus}(e)) further supports the presence of weak magnetic ordering in the title compound.
\\
 To further confirm the presence of subdominant ferromagnetic interactions, isotherm magnetization measurements were performed at several temperatures as shown in Supplementary Fig.~2(b). 
At 0.4 K, a weak but distinct hysteresis at low fields supports the presence of weak ferromagnetic moment below $T_{\rm N}$ (inset of Fig.~\ref{sus}(f)). Above 6 K, however, the linear response indicates dominant antiferromagnetic interactions. \textcolor{black}{
Before proceeding, we note that the presence of subdominant ferromagnetic interactions, the observed ZFC/FC splitting, and the absence of frequency dependence in the $ac$ susceptibility collectively suggest that the non-centrosymmetric chiral crystal structure may give rise to moment canting via Dzyaloshinskii--Moriya interactions (DMIs). This, in turn, could lead to weak canted antiferromagnetic ordering below $T_\mathrm{N}$~\cite{PhysRev.120.91,PhysRevB.109.184432}. To determine precisely the nature of the magnetic ordering, further neutron diffraction studies are required.}
 \\
Pulsed-field magnetization measurements up to  55 T show saturation at ~5 $\mu_{\rm B}$ above 43 T, with no sign of a 1/3 magnetization plateau unlike the classical spin-liquid candidate Na[Mn(HCOO)$_{3}$]  ($S$ = 5/2) with a single trillium lattice  \cite{PhysRevLett.128.177201}. This absence implies that the hypertrillium lattice involves additional interactions beyond the Heisenberg model.
\subsection{Magnetic specific heat}
\begin{figure}[t]
   	\centering
   	\includegraphics[ width=0.5\textwidth]{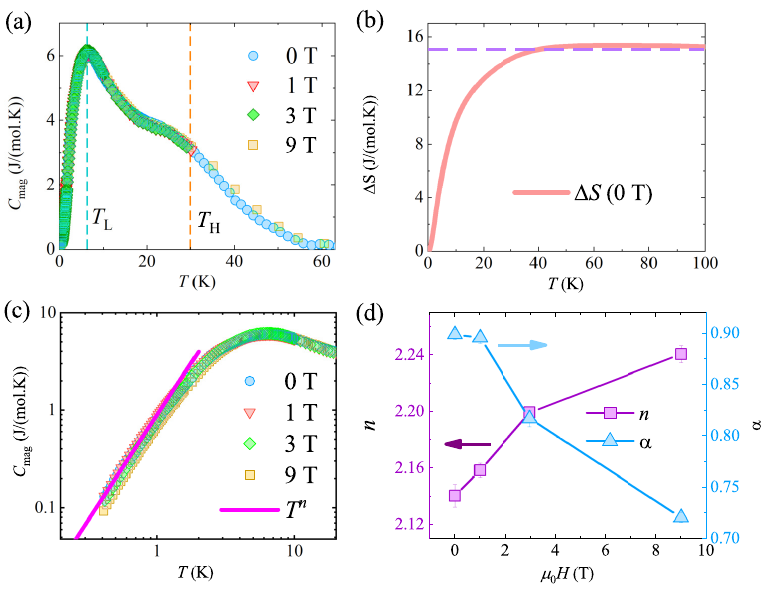}
   	\caption{(a) Temperature dependence of magnetic specific heat ($C_{\rm mag}$) at several magnetic fields.  The vertical  dashed line represents two characteristic temperatures around $T_{\rm L}$ = 6.3 K and $T_{\rm H}$ = 25 K.  (b) Temperature dependence of entropy change at zero-field. \textcolor{black}{The dashed horizontal line indicates the expected entropy, $R$ln$(2S+1)$, for $S$ = 5/2.} (c) $C_{\rm mag}$ as a function of temperature  where the solid red line showing the power-law behavior  ($\approx$ $\alpha T^{n}$). (d) Field dependence of exponent ($n$) and the coefficient ($\alpha$). Error bars represent an uncertainty of one standard deviation. }{\label{HC}}.
   \end{figure}

To gain deeper insights into the magnetic behavior, specific heat measurements were conducted under various magnetic fields.
 Figure~\ref{HC}(a) shows the temperature dependence of magnetic specific heat, $C_{\rm mag}$, obtained after subtracting  lattice contributions at several magnetic fields (see Supplementary Note 3 \cite{sm}) \cite{PhysRevB.63.212408,PhysRevB.110.184402}.  \\ $C_{\rm mag}$
 begins to increase below 70 K, suggesting the growth magnetic correlations well above $\theta_{\rm CW}$. A broad peak around  $T_{\rm H}$ = 25 K signifies that these correlations evolve toward saturation, consistent with the $\chi(T)$ data (Fig.~\ref{sus}(b)).
 In classical spin systems with competing interactions, the spin configuration remains degenerate at finite temperatures, fluctuating among multiple low-energy states before stabilizing at lower temperatures due to subleading terms \cite{PhysRevB.86.205134}. This high-temperature regime, often termed as a classical paramagnetic region, spans from $T_{\rm H}$ to $\sim$ 66 K in our system. Notably, the pronounced anomaly around $T_{\rm L}$ = 6.3 K alludes to the formation of a well-structured spin configuration, potentially driven by subdominant ferromagnetic interactions \cite{Lozano2024}.
   Temperature dependence of  entropy change is shown in Fig.~\ref{HC}(b) at zero-field. The power-law behavior (see below) is extrapolated to zero temperature to compute a total magnetic entropy.  At $T_{\rm L}$, the magnetic entropy  is 0.45$R$ln(6), reflecting partial entropy release, typical of classical systems with a degenerate ground state.  The entropy reaches its full theoretical value, Rln(6), above $\theta_{\rm CW}$. The absence of any anomaly in $C_{\rm mag}$ from weak magnetic ordering around $T_{\rm N}$, observed in $\chi(T)$, highlights its subtle nature. Furthermore, the field-dependent suppression of the weak kink in $\chi(T)$ is reflected in the low-temperature behavior of $C_{\rm mag}$ below 3 K.   \\ 
At $T \ll T_{\rm N}$, $C_{\rm mag}$ shows a power-law dependence, $C_{\rm mag}$ = $\alpha T^{n}$, indicating the presence of gapless excitations, as shown in Fig.~\ref{HC}(c).  The obtained exponent $n$ and coefficient $\alpha$ are plotted as functions of the applied magnetic field in Fig.~\ref{HC}(d). Interestingly, the exponent increases from 2.13(1) to 2.24(1) with increasing magnetic field up to 9 T. The associated  density of excitations, represented by 
$\alpha$, is suppressed with increasing field. \textcolor{black}{
\textcolor{black}{Notably, the near-quadratic temperature dependence of $C_\mathrm{mag}$ suggests the formation of local spin singlets, akin to the behavior observed in the kagome compound SrCr$_{9p}$Ga$_{12-2p}$O$_{19}$ ($p$ = 0.92)~\cite{PhysRevLett.84.2957}}. More importantly, this quadratic behavior is also found in several 3D spin-liquid candidates, regardless of the presence of disorder~\cite{10.1063/5.0096942,PhysRevLett.127.157204}. \textcolor{black}{ Furthermore, our compound shows no signatures of a random-singlet-like state, typically characterized by a power-law behavior $C_\mathrm{mag}/T \propto T^{-n}$ with $n < 1$~\cite{Khatua2022}. Therefore, the potential influence of Fe/Sn site disorder on the ground state remains an open question and warrants further theoretical investigation.}}
\subsection{Electron spin resonance}
\begin{figure}
	\centering
	\includegraphics[ width=0.48\textwidth]{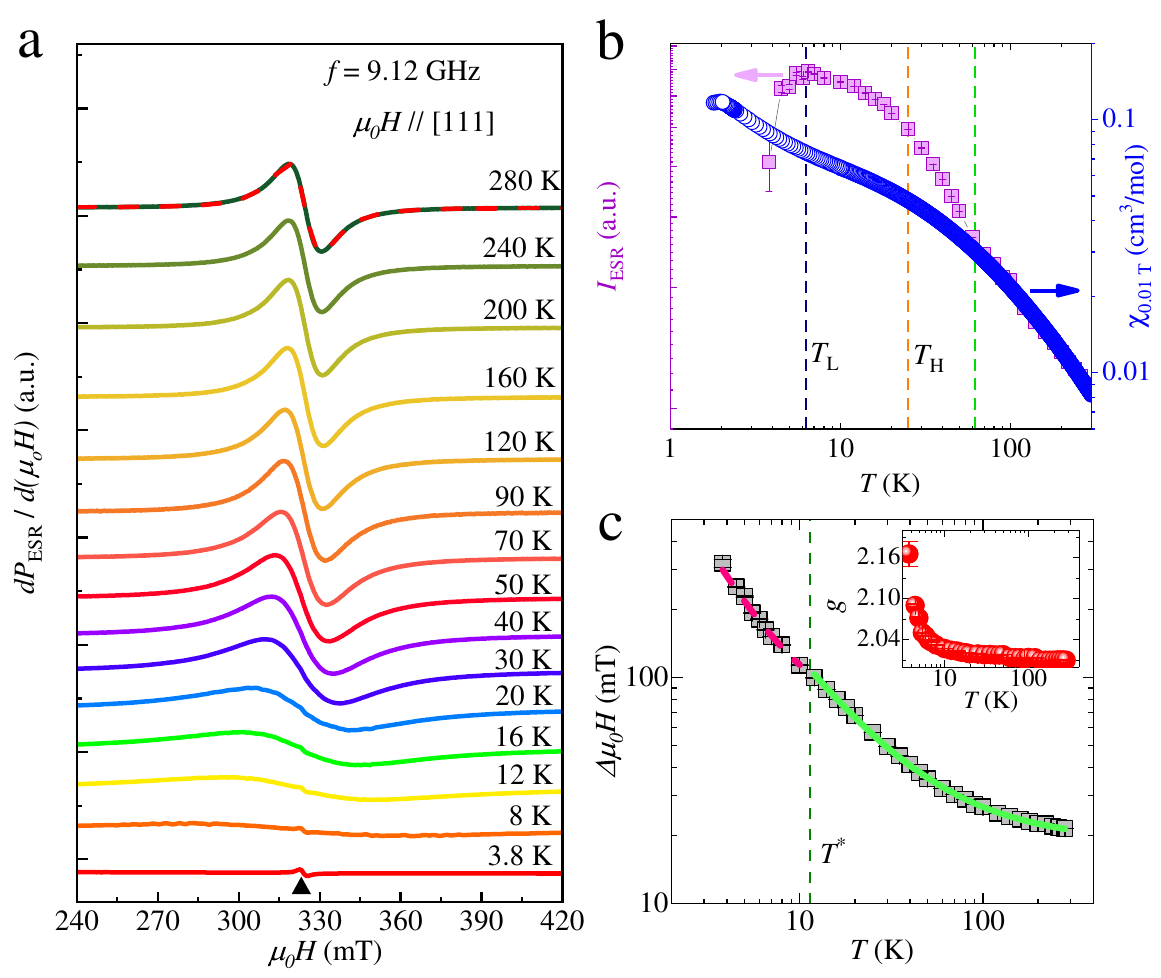}
	\caption{(a) Representative ESR spectra at $f$ = 9.12 GHz at various temperatures with the magnetic field applied to [111] crystal orientation. All spectra are offset for clarity. The dashed red line on the spectrum at $T$ = 280 K represents the fitting using a derivative Lorentzian function.  (b) Temperature dependence of the integrated ESR intensity (left-$y$ axis). The blue data points in right-$y$ axis represent the $\chi(T)$ data at $\mu_{0}H = 0.01$ T from Fig.~\ref{sus}(b). The positions of the dashed vertical lines are detailed in the text. (c)  Temperature dependence of the peak-to-peak linewidth, and the $g$-factor  (in the inset) obtained from the fittings of the ESR spectra shown in (a).  The red dashed and solid green curves in linewidth represent the fit of the data using a power-law function (see the text in detail). The dashed vertical line at $T^{*} \approx 11$ K indicates the crossover between two power-law behaviors. (b) and (c) are log-log plots. Error bars represent an uncertainty of one standard deviation.}{\label{ESR}}. 
\end{figure}

To trace a thermal evolution of spin correlations, X-band ESR measurements were performed down to 3.8 K at $f$ = 9.12 GHz for $\mu_{0}H$ // [111] (Fig.~\ref{ESR}(a)). An exchange-narrowed single resonance line is observed across all temperatures, fitted with a derivative Lorentzian function. As the temperature decreases, the ESR signal  broadens all the way down to 3.8 K. A signal from an experimental artifact is marked by a triangle at the lowest temperature.
The extracted ESR parameters, including integrated intensity, $g$-factor, and linewidth, as a function of temperature, are shown in Fig.~\ref{ESR}(b) and Fig.~\ref{ESR}(c). The integrated intensity follows $\chi(T)$ down to 65 K (green dashed line in Fig.~\ref{ESR}(b)), after which it deviates and increases sharply. Below 7 K, the intensity abruptly decreases. This deviation coincides with the low-$T$ departure from the high-$T$ CW fit, 
suggesting a change in spin correlations around 65 K.\\
The buildup of the internal field is reflected in the $g$-value shift (inset of Fig.~\ref{ESR}(c)). Above 140 K, the $g$-value is 2.010(3), close to the expected 2.002(3) for half-filled Fe$^{3+}$ ($S$ = 5/2), and increases gradually as the temperature decreases. At 3.8 K, the $g$-value rises sharply to 2.16, indicating the development of critical-like spin fluctuations towards 3.8 K. \\
The temperature-dependent evolution of spin correlations is more evident in the linewidth broadening (Fig.~\ref{ESR}(c)). The ESR linewidth increases across the entire temperature range, even up to 280 K. This persistence of linewidth broadening up to $\sim$7$\theta_{\rm CW}$ is typical for low-dimensional frustrated magnets but unusual for 3D frustrated lattices.\\ 
To further analyze this, the ESR linewidth is fitted with a power-law function, $\mu_{0}\Delta H \propto T^{-p}$. The fitting reveals two distinct regimes: $p$ = 1.12 $\pm$ 0.01 down to 10 K and $p$ = 1.5 $\pm$ 0.3 below 10 K. These different exponents indicate a change in spin correlations at the crossover temperature $T^{*}$ = 11 K. As discussed, the spin correlations between $T^{*}$ and $T$ = 298 K are attributed to dominant antiferromagnetic Heisenberg interactions, while below $T^{*}$, a subleading term stabilizes a well-defined correlated spin configuration, accompanied by a slowdown of spin fluctuations \cite{PhysRevB.93.174402}. 
We note that two-stage linewidth broadening is also observed in other frustrated magnets, including the trillium lattice compound K$_{2}$CrTi(PO$_{4}$)$_{3}$ \cite{PhysRevB.105.094439,PhysRevB.109.184432,PhysRevB.93.174402,PhysRevB.95.184430}. 
\subsection{Muon spin relaxation}\label{frr}
 To gain additional insight into spin dynamics in KFSPO, ZF and  LF $\mu$SR measurements were performed over wide temperature and field ranges. The  implanted 100\text{\%} spin-polarized muons localize at an interstitial lattice site, typically  1 {\AA} from oxygen ions in KFSPO \cite{le2011muon}.\\  \begin{figure*}
 	\centering
 	\includegraphics[ width=\textwidth]{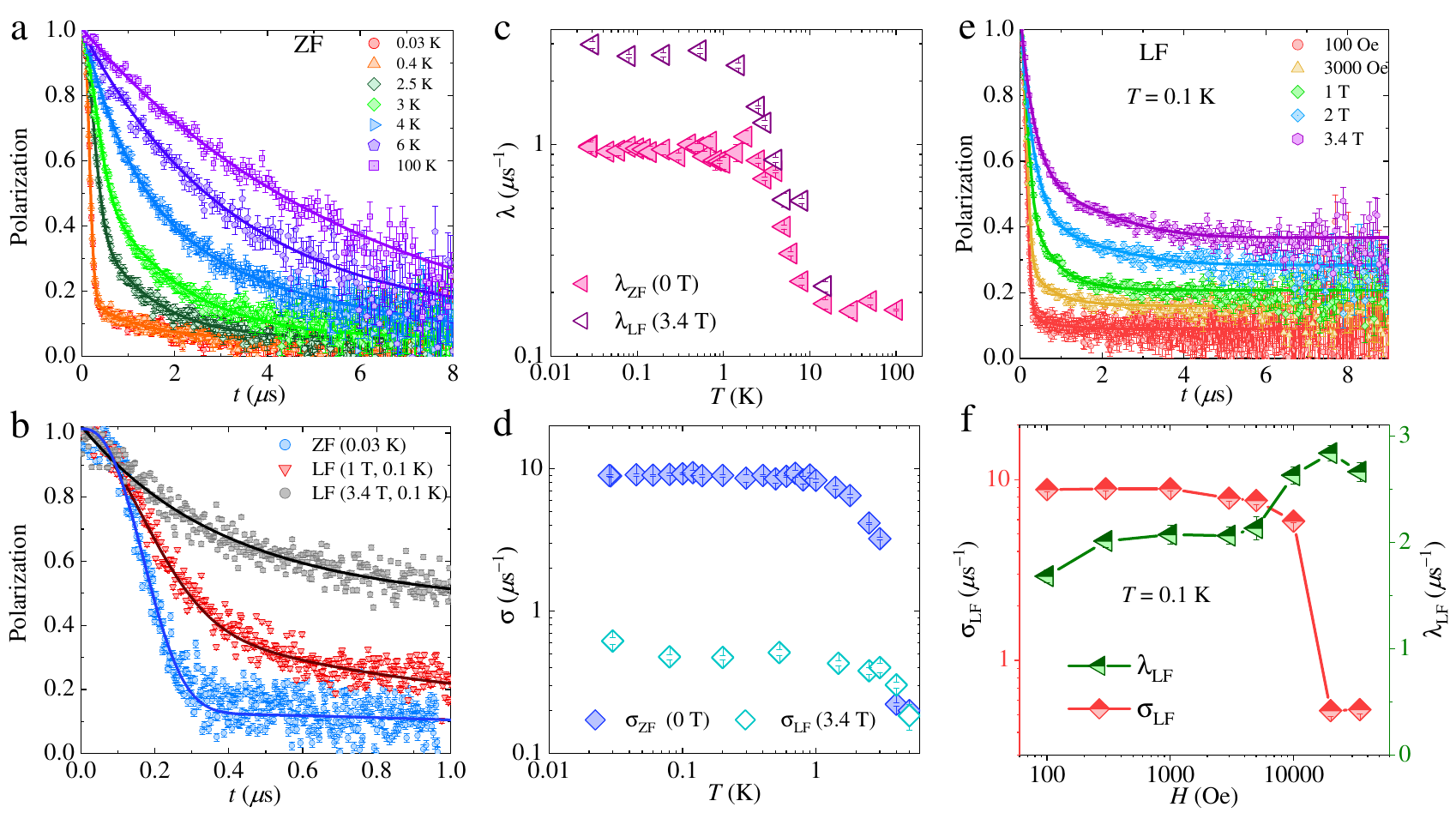}
 	\caption{(a) Time evolution of normalized muon spin polarization in ZF for a few representative temperatures over a long-time scale.  (b) Time evolution of normalized muon spin polarization in ZF at $T$ = 0.03 K and under two LFs at  $T$ = 0.1 K at short times. (c)  Temperature dependence of the exponential muon spin relaxation rates at late times in ZF and LF of $\mu_{0}H$ = 3.4 T. (d) Temperature dependence of the Gaussian relaxation rates at early times in ZF and LF of $\mu_{0}H$ = 3.4 T. (e) Field dependence of the normalized muon spin polarization at $T$ = 0.1 K.  (f) Gaussian (left\textit{ y}-axis) and exponential (right \textit{y}-axis) muon spin relaxation rate as a function of LF at 0.1 K. Error bars represent an uncertainty of one standard deviation.
 	}{\label{musr}}.
 \end{figure*} 
In the polycrystalline samples with long-range magnetic order, the muon spin polarization typically exhibits coherent oscillations corresponding to the 2/3 transverse component and a non-oscillating 1/3 longitudinal component. As shown in Fig.~\ref{musr}(a), the time evolution of the muon spin polarization down to $T$ = 30 mK reveals neither coherent oscillations indicative of static magnetic order, nor the typical ``1/3'' tail associated with the randomly oriented static internal fields (spin-freezing) in KFSPO \cite{PhysRevB.31.546}. On the other hand, in the fast fluctuation limit—where all three spatial components become equivalent—one would expect the absence of the characteristic ``1/3'' tail. In KFSPO, as the temperature decreases below $T_{\rm L}$, the polarization shows a gradual development of rapid decay at early times ($t \ll 0.2\ \mu$s), (Fig.~\ref{musr}(b)) indicating  muon senses a static distribution of internal magnetic fields over the muon lifetime, while the remaining relaxation persisting to late times suggests the presence of fluctuating electronic moments. This scenario implies that KFSPO resides in an intermediate regime between the static and fast fluctuation limits, consistent with the presence of weak magnetic order. \textcolor{black}{ Furthermore, as discussed in the following section, the observed muon spin polarization differs from that often seen in 3D spin-liquid candidates where static moments from spin freezing coexist with dynamic moments~\cite{Sibille2017,Cai2018}}.\\ 
For a quantitative analysis, various models were tested, and the function $P(t)_{\rm ZF} = \exp[-\lambda_{\rm ZF} t ] +$ Bg (solid line in Fig.~\ref{musr}(a)) was found to best describe the ZF spectra down to $T_{\rm L}$, where Bg represents a constant background from muons stopping outside the sample. This Lorentzian form of muon spin polarization captures the cooperative dynamics of fluctuating local electronic moments,  which give rise to the muon spin relaxation rate $\lambda_{\rm ZF}$. However, for $T < T_{\rm L}$, the ZF spectra down to 30 mK require the polarization function (solid line in Fig.~\ref{musr}(a)) $P(t)_{\rm ZF} = f_{\rm ZF}\rm{exp}[-\lambda_{\rm ZF} \textit{t}] + (1-\textit{f}_{\rm ZF})\exp[-\frac{1}{2}(\sigma_{ZF} \textit{t})^2]$ + Bg,  where the second term, which dominates at early times and reflects a Gaussian distribution of quasistatic internal fields at the muon site, becomes necessary to account for the low-temperature behavior of the polarization (see Fig.~\ref{musr}(b)). Here, $f_{\rm ZF}$ and $(1 - f_{\rm ZF})$ represent the fractions of dynamic and quasistatic components of the signal, respectively, while $\sigma_{\rm ZF}$ corresponds to the muon spin relaxation rate associated to the fast relaxing static component. 
  \\
The temperature dependence of  obtained $\lambda_{\rm ZF}$ is shown in Fig.~\ref{musr}(c) in ZF. For $T > T^{*}$ = 11 K, $\lambda_{\rm ZF}$ remains constant originating from the exchange fluctuations of electronic spins of Fe$^{3+}$ ions in the motional narrowing regime. Using the mean-field approximation for  average exchange interaction $ J = 3k_{\rm B} \theta_{\rm CW}/2zS(S+1) $  with the nearest-neighbor coordination number $ z = 6 $ and $ S = 5/2 $, the exchange fluctuation rate is estimated as  $\nu = \sqrt{z} J S/{\hbar} = 9.78 \times 10^{11} \text{ s}^{-1}$ \cite{PhysRevLett.73.3306}. 
This gives the internal field distribution width \mbox{$\Delta/\gamma_{\mu}$ =  $\sqrt{\nu \lambda_{\rm ZF}/2}/\gamma_{\mu} \approx  3$ kG} at high temperatures,  where $\lambda_{\rm ZF} = 0.165$ $\mu \text{s}^{-1}$ ($T\gg T^{*}$) and the muon gyromagnetic ratio $\gamma_{\mu}$ = 2$\pi$ $\times$ 135.5 MHz/T. As the temperature decreases below $T^{*}$, $\lambda_{\rm ZF}$ increases by nearly a factor of six, attributed to the slowing down of spin fluctuations. Furthermore, the $T$-independent plateau behavior observed at low temperatures indicates the presence of persistent spin dynamics, \textcolor{black}{a hallmark of frustrated magnets with strong spin fluctuations}~\cite{PhysRevLett.73.3306}. Persistent spin dynamics accommodated by the two components of the muon spin relaxation rate, are also observed in other trillium lattice systems despite undergoing symmetry-breaking phase transitions \cite{PhysRevLett.127.157204,PhysRevB.109.184432}. \\
The extracted Gaussian relaxation rate, which originates from quasistatic local moments, is shown in Fig.~\ref{musr}(d) as a function of temperature. It exhibits a sudden increase by several orders of magnitude with a slight temperature rise below $T_{\rm L}$, followed by a tendency toward saturation below 1 K. The temperature independence of $\sigma_{\rm ZF}$ below 1 K suggests that the system enters a quasistatic regime characterized by slow spin dynamics. 
The temperature dependence of $f_{\rm ZF}$ and $(1 - f_{\rm ZF})$, shown in Supplementary Fig.~4(a), reflects the fractions of dynamic and static components, respectively. Notably, both components remain nearly temperature-independent below 1 K with $\approx$ 80\text{\%} quasistatic and $\approx$ 18 \text{\%} dynamic signal fractions, indicating the coexistence of quasistatic and fluctuating local electronic moments in the ground state. However, a distinct increase of  $f_{\rm ZF}$  accompanied by a corresponding decrease of $f_{\rm ZF}$ is observed down to $T_{\rm L}$. Above $T_{\rm L}$, the dynamic fraction persists to higher temperatures and closely tracks the crossover temperature $T^{*}$. \textcolor{black}{ 
It is worth noting that at low temperatures the $\sim$ 18\% exponential relaxation component is primarily attributed to weakly ordered moments that remain dynamic. The remaining $\sim$ 80\%, exhibiting anomalous Gaussian-like muon spin polarization, likely \textcolor{black}{reflects characteristics common to strongly correlated magnets hosting  spin singlets}~\cite{PhysRevLett.73.3306,PhysRevLett.127.157204}. In the trillium lattice K$_2$Ni$_2$(SO$_4$)$_3$, a similar exponential component has been attributed to partial homogeneous ordering \cite{PhysRevLett.127.157204}, while the remaining relaxation arises from sporadic unpaired-spin excitations, consistent with our observation of an undecoupled Gaussian relaxation in both ZF and LF data (discussed below). Compared to the $S=1$ trillium lattice system K$_2$Ni$_2$(SO$_4$)$_3$~\cite{PhysRevLett.127.157204}, our $S=5/2$ system exhibits more classical spin behavior, characterized by a stronger growth of the static fraction below $T_\mathrm{L}$, likely due to sporadic unpaired-spin excitations, as suggested by the decoupling experiments described below~ \cite{PhysRevLett.127.157204,PhysRevB.109.184432,PhysRevLett.73.3306}.}	
\\
In order to find the intrinsic origin of Gaussian distribution of local internal fields  at low temperatures, $\mu$SR measurements were conducted at $T= 0.1 $ K under several LFs which decouples the polarization from  purely static and random local fields. As shown in Fig.~\ref{musr}(e),  even when applying LFs exceeding the expected local field from ZF data ($\Delta/\gamma_{\mu} \approx 3$ kG), the decoupling of the muon spin polarization remains minimal.
 This  weakness of the field dependence of polarization suggests that the Gaussian distribution of local fields manifests a distinct fingerprint in the present correlated system.  Furthermore, if the Gaussian distribution of  local fields were static at low temperatures, full polarization would be recovered in a LF of 3.4 T (Fig.~\ref{musr}(e)).
The strong residual ``undecouplable Gaussian'' line shape at early times can be attributed to the presence  spin excitations (e.g., spinons) which contribute to Gaussian shape muon relaxation on the background of a quasistatic frozen state \cite{PhysRevLett.73.3306}. A similar behavior has been observed in other spin liquid candidates including the double trillium lattice compound K$_2$Ni$_2$(SO$_4$)$_3$ \cite{PhysRevLett.127.157204} and  the kagome lattice SrCr$_{8}$Ga$_{4}$O$_{19}$ \cite{PhysRevLett.73.3306}.\\
To further confirm the robustness of the  significant spin fluctuations even under a high magnetic field, $\mu$SR measurements were conducted at a LF of 3.4 T in the $T$-range 30 mK $\leq T \leq$ 15 K (see Supplementary Fig.~4(b)).
 The  $\mu$SR data in LF were modeled using the function
$P(t)_{\rm LF} = f_{\rm LF}\rm{exp}[-\lambda_{\rm LF} \textit{t} ] + (1-\textit{f}_{\rm LF})\exp[-\frac{1}{2}(\sigma_{LF} \textit{t})^2]$ + B$_{\rm g}$.
 The parameter $f_{\rm LF}$ (1$-f_{\rm LF}$) corresponds to the fraction of the dynamic(quasistatic)-relaxing
component. Its value remains nearly temperature-independent below $T_{\rm L}$, with approximately 80$\text{\%}$ dynamic and 20\text{\%} static contributions (see Supplementary Fig.~4(c)).  The obtained $\lambda_{\rm LF}$ and $\sigma_{\rm LF}$ relaxation rates in a LF of $\mu_{0}H$ = 3.4 T  are compared
with those obtained from the ZF data in Fig.~\ref{musr}(c) and Fig.~\ref{musr}(d). The temperature dependence of both relaxation rates at 3.4 T reveals the presence of persistent spin dynamics within a weak quasistatic regime.\\ Interestingly, in a LF of 3.4 T particularly below $T_{\rm L}$, the quasistatic component decreases by 75\text{\%} compared to ZF, and the corresponding relaxation rate drops sharply  by 93 \text{\%}, indicating the suppression of low-energy spin excitations that govern the muon spin relaxation. This is further corroborated by the field dependence of the parameter $\alpha$ (Fig.~\ref{HC}(d)), which reflects the density of such excitations. On the other hand, the dynamic component increases nearly fivefold, and its relaxation rate rises by almost three orders of magnitude, suggesting the decoupling of static local fields—consistent with the field-induced suppression of weak magnetic order and field-resilient spin dynamics, as evidenced by the enhanced Lorentzian fraction and rate.
 \\ 
Figure.~\ref{musr}(f) shows the LF dependence of $\sigma_{\rm LF}$ (left $y$-axis) and $\lambda_{\rm LF}$ (right $x$-axis),  highlighting the suppression of low-energy excitations—evident from the reduced Gaussian relaxation rate and enhanced dynamic relaxation rate due to the decoupling of quasistatic fields. This scenario can be correlated with the field dependence of the coefficient $\alpha$ and the power-law exponent $n$ in the magnetic specific heat expression, $C_{\rm mag} \approx \alpha T^{n}$ at low temperatures.
	\section{DISCUSSION}
\textcolor{black}{ We first address the weak magnetic ordering at $T_\mathrm{N} \approx 2$~K, seen from subtle kinks in both $dc$ and $ac$ magnetic susceptibility. We recall that the non-centrosymmetric chiral cubic space group $P2_1 3$ lacks inversion symmetry but permits DMIs, which can induce weak ferromagnetism~\cite{PhysRev.120.91}. For example, in the trillium lattice compound Cs$_2$Fe$_2$(MoO$_4$)$_3$ ($T_\mathrm{N} \approx 1$~K), DMIs are thought to contribute to the presence of subdominant ferromagnetic exchange~\cite{Kub2024}. Similarly, trillium lattices in $P2_1 3$ often host ferromagnetic interactions in addition to dominant antiferromagnetic ones~\cite{PhysRevB.109.184432,PhysRevLett.106.156603,BATTLE198616}. Given that KFSPO shares this symmetry, weak ferromagnetism driven by DMIs is expected, consistent with the canted antiferromagnetic order seen in the related compound K$_{2}$CrTi(PO$_4$)$_3$~\cite{PhysRevB.109.184432}.	However, unlike magnetic susceptibility, magnetic specific heat and $\mu$SR do not clearly detect this ordering, highlighting its subtlety. Such hidden order is often reported in trillium compounds; for instance, K$_2$CrTi(PO$_4$)$_3$ shows no ordering in susceptibility but clear signatures in specific heat~\cite{PhysRevB.109.184432}, while in K$_2$Ni$_2$(SO$_4$)$_3$, specific heat reveals ordering that $\mu$SR does not clearly capture~\cite{PhysRevLett.127.157204}.
\\ Despite the presence of weak magnetic order, it is observed that strong spin fluctuations, which may arise from the hypertrillium spin topology, can lead   persistent spin dynamics without a clear signature of magnetic ordering as observed in $\mu$SR experiments for K$_2$Ni$_2$(SO$_4$)$_3$ \cite{PhysRevLett.127.157204}. A similar scenario is expected in the titled compound KFSPO, where weak magnetic ordering may lead to a fraction of quasistatic fast relaxation superimposed on a dominant undecoupled Gaussian relaxation associated with non-trivial spin excitations, along with additional relaxation from persistent spin dynamics.
   This Gaussianlike relaxation from low-energy excitations is also observed in K$_2$Ni$_2$(SO$_4$)$_3$, indicating a common feature of the hypertrillium spin topology that is maintained despite anti-site disorder in KFSPO. Most importantly, if the Gaussian relaxation originates entirely from disorder-induced quasistatic moments, this should be reflected in our decoupling experiments as a saturation of the full polarization at fields  3.4\,T $> \Delta/\gamma_{\mu}$. Therefore, the absence of full recovery of muon spin polarization in LF-$\mu$SR experiments provides concrete evidence for disorder-induced smearing of the LRO state in KFSPO. An alternative scenario of disorder-induced quasistatic magnetic field is often identified through Gaussian-broadened relaxation of the muon spin polarization function, as observed in $\alpha$-Ru$_{x}$Ir$_{1-x}$Cl$_{3}$ \cite{PhysRevB.98.014407}  and K$_2$CrTi(PO$_4$)$_3$ \cite{PhysRevB.109.184432}. The anomalous LF dependence of the muon spin polarization in KFSPO may suggest the presence of \textcolor{black}{intriguing spin fluctuations}, similar to that observed in the $S$ = 3/2 SrCr$_{8}$Ga$_{4}$O$_{19}$ system \cite{PhysRevLett.73.3306}.
      \\  
     A close comparison between the two isostructural $S=5/2$ trillium compounds, KFSPO and KSrFe$_2$(PO$_4$)$_3$~\cite{10.1063/5.0096942}---where KFSPO exhibits magnetic site dilution (Fe/Sn) and KSrFe$_2$(PO$_4$)$_3$ displays non-magnetic site disorder (Sr/K)---reveals that the key difference lies in the temperature at which the ZFC/FC splitting occurs. In contrast, the specific heat behavior is remarkably similar in both systems. This suggests that magnetic site dilution primarily affects the weak ferromagnetic interactions while preserving the essential spin dynamics associated with the trillium lattice topology. Nevertheless, given the substantial magnetic site randomness in KFSPO, a natural question arises regarding the possibility of a random-singlet state~\cite{PhysRevB.111.014409,PhysRevLett.126.037201,shimokawah}, similar to those proposed in other disordered frustrated magnets, such as the $S=1/2$ systems Sr$_2$CuTe$_{1-x}$W$_x$O$_6$~\cite{PhysRevLett.126.037201} and Li$_4$CuTeO$_6$~\cite{Khatua2022}, and  the $S=5/2$ system Lu$_3$Sb$_3$Mn$_2$O$_{14}$~\cite{PhysRevB.107.214404}.     
 \\ Experimental signatures of the random singlet state manifest as power-law behaviors in physical observables such as magnetic susceptibility ($\chi \propto T^{-\alpha}$), specific heat ($C_{\mathrm{mag}}/T \propto T^{-\alpha}$), ESR linewidth ($\Delta H \propto T^{-\alpha}$), and muon spin relaxation rate ($\lambda_{\mathrm{ZF}} \propto T^{-\alpha}$) \cite{PhysRevLett.126.037201,Khatua2022,PhysRevB.107.214404}. The exponent $\alpha$ (0 $<$ $\alpha$ $<$ 1) reflects the probability distribution of random exchange interactions, described by $P(J) \propto J^{-\alpha}$. The absence of a similar power-law dependence in KFSPO indicates that its ground state deviates from the random singlet state.  
 The absence of a characteristic power-law dependence and its scaling in KFSPO suggests that its ground state deviates from a random-singlet scenario. There is no direct calculation for the percolation threshold 
 of a trillium lattice.  The lack of random-singlet phenomenology may be attributed to the relatively low percolation threshold of the trillium lattice. It is highly likely that the Fe occupancy on the two interpenetrating trillium sublattices exceeds this threshold, thereby preserving long-range magnetic connectivity despite Fe/Sn site disorder. Notably, KFSPO shows closer resemblance to the spin dynamics observed in the $S=3/2$ system SrCr$_8$Ga$_4$O$_{19}$~\cite{PhysRevLett.73.3306,PhysRevLett.84.2957,PhysRevB.58.12049}, and exhibits short-range spin correlations above $T_\mathrm{N}$, similar to those reported in the $S=5/2$ classical spin system Li$_9$Fe$_3$(P$_2$O$_7$)$_3$(PO$_4$)$_2$~\cite{PhysRevLett.127.157202}. Although the title compound displays experimental features characteristic of \textcolor{black}{strong spin fluctuations}, subleading interactions and disorder appear to stabilize weak magnetic ordering and introduce inhomogeneity in the spin dynamics and the ground state.}
\begin{center}
\section{CONCLUSION}
\end{center}
In summary, through a combination of bulk and local probe experiments, we investigate the ground-state properties of a 3D trillium lattice compound  K$_{2}$FeSn(PO$_{4}$)$_{3}$, where Fe$^{3+}$ ions form a chiral network of corner-sharing tetrahedra, known as a hypertrillium lattice. The two-step evolution of magnetic correlations, as revealed by ESR linewidth, coupled with two distinct anomalies in specific heat separated by $T^{*}\approx$ 11 K, points to a complex interplay between dominant antiferromagnetic Heisenberg interactions and subleading interaction terms in this system. Furthermore, the weak magnetic ordering at $T_{\rm N} \approx 2$ K is revealed only through \textit{dc }and \textit{ac} magnetic susceptibility measurements, suggesting its subtle nature. This weak ordering is suppressed under an applied field $\mu_{0}H \geq 2$ T, corroborated by the weak field dependence of the magnetic specific heat. Despite the symmetry-breaking transition, $\mu$SR results reveal dominant persistent spin dynamics, \textcolor{black}{pointing to a ground state in which dynamically fluctuating spins coexist with weak magnetic order.}
 Our experimental results underscore the importance of further studies, particularly using inelastic neutron scattering, to clarify the nature of the subleading interactions.\\
 \section{DATA AVAILABILITY}
 \textcolor{black}{The $\mu$SR data that support the findings of this
 paper are openly available \cite{psidatabase}.}
 All other data, including magnetic susceptibility, magnetization, specific heat, and ESR, are available upon reasonable request.
\section{ ACKNOWLEDGMENTS}
	The work at
SKKU was supported by the National Research Foundation (NRF) of Korea (Grants No. RS-2023-00209121 and
No. 2020R1A5A1016518). E.M. is
supported by the Canada Research Chairs program, the
Natural Science and Engineering Research Council of
Canada, and the Canadian Foundation for Innovation.
 This work
was supported by HLD-HZDR, member of the European
Magnetic Field Laboratory (EMFL). R. S. acknowledges the financial support provided by the Ministry of Science and Technology in Taiwan under Project No. NSTC-113-2124-M-001-003 and No. NSTC-113-2112M001-045-MY3,  as well as support from Academia Sinica for the budget of AS-iMATE11312. financial support from the Center of Atomic Initiative for New Materials (AIMat), National Taiwan University,  under Project No. 113L900801. This work is partially based on experiments performed at the Swiss Muon
Source SµS, Paul Scherrer Institute, Villigen, Switzerland.
\bibliography{Trillium.bib}

\begin{thebibliography}{81}%
\makeatletter
\providecommand \@ifxundefined [1]{%
 \@ifx{#1\undefined}
}%
\providecommand \@ifnum [1]{%
 \ifnum #1\expandafter \@firstoftwo
 \else \expandafter \@secondoftwo
 \fi
}%
\providecommand \@ifx [1]{%
 \ifx #1\expandafter \@firstoftwo
 \else \expandafter \@secondoftwo
 \fi
}%
\providecommand \natexlab [1]{#1}%
\providecommand \enquote  [1]{``#1''}%
\providecommand \bibnamefont  [1]{#1}%
\providecommand \bibfnamefont [1]{#1}%
\providecommand \citenamefont [1]{#1}%
\providecommand \href@noop [0]{\@secondoftwo}%
\providecommand \href [0]{\begingroup \@sanitize@url \@href}%
\providecommand \@href[1]{\@@startlink{#1}\@@href}%
\providecommand \@@href[1]{\endgroup#1\@@endlink}%
\providecommand \@sanitize@url [0]{\catcode `\\12\catcode `\$12\catcode
  `\&12\catcode `\#12\catcode `\^12\catcode `\_12\catcode `\%12\relax}%
\providecommand \@@startlink[1]{}%
\providecommand \@@endlink[0]{}%
\providecommand \url  [0]{\begingroup\@sanitize@url \@url }%
\providecommand \@url [1]{\endgroup\@href {#1}{\urlprefix }}%
\providecommand \urlprefix  [0]{URL }%
\providecommand \Eprint [0]{\href }%
\providecommand \doibase [0]{https://doi.org/}%
\providecommand \selectlanguage [0]{\@gobble}%
\providecommand \bibinfo  [0]{\@secondoftwo}%
\providecommand \bibfield  [0]{\@secondoftwo}%
\providecommand \translation [1]{[#1]}%
\providecommand \BibitemOpen [0]{}%
\providecommand \bibitemStop [0]{}%
\providecommand \bibitemNoStop [0]{.\EOS\space}%
\providecommand \EOS [0]{\spacefactor3000\relax}%
\providecommand \BibitemShut  [1]{\csname bibitem#1\endcsname}%
\let\auto@bib@innerbib\@empty
\bibitem [{\citenamefont {Balents}(2010)}]{Balents2010}%
  \BibitemOpen
  \bibfield  {author} {\bibinfo {author} {\bibfnamefont {L.}~\bibnamefont
  {Balents}},\ }\bibfield  {title} {\bibinfo {title} {Spin liquids in
  frustrated magnets},\ }\href {https://doi.org/10.1038/nature08917} {\bibfield
   {journal} {\bibinfo  {journal} {Nature}\ }\textbf {\bibinfo {volume}
  {464}},\ \bibinfo {pages} {199} (\bibinfo {year} {2010})}\BibitemShut
  {NoStop}%
\bibitem [{\citenamefont {Savary}\ and\ \citenamefont
  {Balents}(2016)}]{Savary_2016}%
  \BibitemOpen
  \bibfield  {author} {\bibinfo {author} {\bibfnamefont {L.}~\bibnamefont
  {Savary}}\ and\ \bibinfo {author} {\bibfnamefont {L.}~\bibnamefont
  {Balents}},\ }\bibfield  {title} {\bibinfo {title} {Quantum spin liquids: a
  review},\ }\href {https://doi.org/10.1088/0034-4885/80/1/016502} {\bibfield
  {journal} {\bibinfo  {journal} {Rep. Prog. Phys.}\ }\textbf {\bibinfo
  {volume} {80}},\ \bibinfo {pages} {016502} (\bibinfo {year}
  {2016})}\BibitemShut {NoStop}%
\bibitem [{\citenamefont {Henley}(2010)}]{annurev38}%
  \BibitemOpen
  \bibfield  {author} {\bibinfo {author} {\bibfnamefont {C.~L.}\ \bibnamefont
  {Henley}},\ }\bibfield  {title} {\bibinfo {title} {The “coulomb phase” in
  frustrated systems},\ }\href
  {https://doi.org/https://doi.org/10.1146/annurev-conmatphys-070909-104138}
  {\bibfield  {journal} {\bibinfo  {journal} {Annu. Rev. Condens. Matter
  Phys.}\ }\textbf {\bibinfo {volume} {1}},\ \bibinfo {pages} {179} (\bibinfo
  {year} {2010})}\BibitemShut {NoStop}%
\bibitem [{\citenamefont {Lau}\ \emph {et~al.}(2006)\citenamefont {Lau},
  \citenamefont {Freitas}, \citenamefont {Ueland}, \citenamefont {Muegge},
  \citenamefont {Duncan}, \citenamefont {Schiffer},\ and\ \citenamefont
  {Cava}}]{Lau2006}%
  \BibitemOpen
  \bibfield  {author} {\bibinfo {author} {\bibfnamefont {G.~C.}\ \bibnamefont
  {Lau}}, \bibinfo {author} {\bibfnamefont {R.~S.}\ \bibnamefont {Freitas}},
  \bibinfo {author} {\bibfnamefont {B.~G.}\ \bibnamefont {Ueland}}, \bibinfo
  {author} {\bibfnamefont {B.~D.}\ \bibnamefont {Muegge}}, \bibinfo {author}
  {\bibfnamefont {E.~L.}\ \bibnamefont {Duncan}}, \bibinfo {author}
  {\bibfnamefont {P.}~\bibnamefont {Schiffer}},\ and\ \bibinfo {author}
  {\bibfnamefont {R.~J.}\ \bibnamefont {Cava}},\ }\bibfield  {title} {\bibinfo
  {title} {Zero-point entropy in stuffed spin-ice},\ }\href
  {https://doi.org/10.1038/nphys270} {\bibfield  {journal} {\bibinfo  {journal}
  {Nat. Phys.}\ }\textbf {\bibinfo {volume} {2}},\ \bibinfo {pages} {249}
  (\bibinfo {year} {2006})}\BibitemShut {NoStop}%
\bibitem [{\citenamefont {Bacciconi}\ \emph {et~al.}(2024)\citenamefont
  {Bacciconi}, \citenamefont {Xavier}, \citenamefont {Carusotto}, \citenamefont
  {Chanda},\ and\ \citenamefont
  {Dalmonte}}]{bacciconi2024theoryfractionalquantumhall}%
  \BibitemOpen
  \bibfield  {author} {\bibinfo {author} {\bibfnamefont {Z.}~\bibnamefont
  {Bacciconi}}, \bibinfo {author} {\bibfnamefont {H.}~\bibnamefont {Xavier}},
  \bibinfo {author} {\bibfnamefont {I.}~\bibnamefont {Carusotto}}, \bibinfo
  {author} {\bibfnamefont {T.}~\bibnamefont {Chanda}},\ and\ \bibinfo {author}
  {\bibfnamefont {M.}~\bibnamefont {Dalmonte}},\ }\href
  {https://arxiv.org/abs/2405.12292} {\bibinfo {title} {Theory of fractional
  quantum \text{H}all liquids coupled to quantum light and emergent
  graviton-polaritons}} (\bibinfo {year} {2024}),\ \Eprint
  {https://arxiv.org/abs/2405.12292} {arXiv:2405.12292 [cond-mat.mes-hall]}
  \BibitemShut {NoStop}%
\bibitem [{\citenamefont {Khatua}\ \emph {et~al.}(2023)\citenamefont {Khatua},
  \citenamefont {Sana}, \citenamefont {Zorko}, \citenamefont {Gomilšek},
  \citenamefont {Sethupathi}, \citenamefont {Rao}, \citenamefont {Baenitz},
  \citenamefont {Schmidt},\ and\ \citenamefont {Khuntia}}]{KHATUA20231}%
  \BibitemOpen
  \bibfield  {author} {\bibinfo {author} {\bibfnamefont {J.}~\bibnamefont
  {Khatua}}, \bibinfo {author} {\bibfnamefont {B.}~\bibnamefont {Sana}},
  \bibinfo {author} {\bibfnamefont {A.}~\bibnamefont {Zorko}}, \bibinfo
  {author} {\bibfnamefont {M.}~\bibnamefont {Gomilšek}}, \bibinfo {author}
  {\bibfnamefont {K.}~\bibnamefont {Sethupathi}}, \bibinfo {author}
  {\bibfnamefont {M.~R.}\ \bibnamefont {Rao}}, \bibinfo {author} {\bibfnamefont
  {M.}~\bibnamefont {Baenitz}}, \bibinfo {author} {\bibfnamefont
  {B.}~\bibnamefont {Schmidt}},\ and\ \bibinfo {author} {\bibfnamefont
  {P.}~\bibnamefont {Khuntia}},\ }\bibfield  {title} {\bibinfo {title}
  {Experimental signatures of quantum and topological states in frustrated
  magnetism},\ }\href
  {https://doi.org/https://doi.org/10.1016/j.physrep.2023.09.008} {\bibfield
  {journal} {\bibinfo  {journal} {Phys. Rep.}\ }\textbf {\bibinfo {volume}
  {1041}},\ \bibinfo {pages} {1} (\bibinfo {year} {2023})}\BibitemShut
  {NoStop}%
\bibitem [{\citenamefont {Takagi}\ \emph {et~al.}(2019)\citenamefont {Takagi},
  \citenamefont {Takayama}, \citenamefont {Jackeli}, \citenamefont
  {Khaliullin},\ and\ \citenamefont {Nagler}}]{Takagi2019}%
  \BibitemOpen
  \bibfield  {author} {\bibinfo {author} {\bibfnamefont {H.}~\bibnamefont
  {Takagi}}, \bibinfo {author} {\bibfnamefont {T.}~\bibnamefont {Takayama}},
  \bibinfo {author} {\bibfnamefont {G.}~\bibnamefont {Jackeli}}, \bibinfo
  {author} {\bibfnamefont {G.}~\bibnamefont {Khaliullin}},\ and\ \bibinfo
  {author} {\bibfnamefont {S.~E.}\ \bibnamefont {Nagler}},\ }\bibfield  {title}
  {\bibinfo {title} {Concept and realization of {K}itaev quantum spin
  liquids},\ }\href {https://doi.org/10.1038/s42254-019-0038-2} {\bibfield
  {journal} {\bibinfo  {journal} {Nat. Rev. Phys.}\ }\textbf {\bibinfo {volume}
  {1}},\ \bibinfo {pages} {264} (\bibinfo {year} {2019})}\BibitemShut {NoStop}%
\bibitem [{\citenamefont {Broholm}\ \emph {et~al.}(2020)\citenamefont
  {Broholm}, \citenamefont {Cava}, \citenamefont {Kivelson}, \citenamefont
  {Nocera}, \citenamefont {Norman},\ and\ \citenamefont
  {Senthil}}]{scienceaay0668}%
  \BibitemOpen
  \bibfield  {author} {\bibinfo {author} {\bibfnamefont {C.}~\bibnamefont
  {Broholm}}, \bibinfo {author} {\bibfnamefont {R.~J.}\ \bibnamefont {Cava}},
  \bibinfo {author} {\bibfnamefont {S.~A.}\ \bibnamefont {Kivelson}}, \bibinfo
  {author} {\bibfnamefont {D.~G.}\ \bibnamefont {Nocera}}, \bibinfo {author}
  {\bibfnamefont {M.~R.}\ \bibnamefont {Norman}},\ and\ \bibinfo {author}
  {\bibfnamefont {T.}~\bibnamefont {Senthil}},\ }\bibfield  {title} {\bibinfo
  {title} {Quantum spin liquids},\ }\href
  {https://doi.org/10.1126/science.aay0668} {\bibfield  {journal} {\bibinfo
  {journal} {Science}\ }\textbf {\bibinfo {volume} {367}},\ \bibinfo {pages}
  {eaay0668} (\bibinfo {year} {2020})}\BibitemShut {NoStop}%
\bibitem [{\citenamefont {Nasu}(2023)}]{10.1093/ptep/ptad115}%
  \BibitemOpen
  \bibfield  {author} {\bibinfo {author} {\bibfnamefont {J.}~\bibnamefont
  {Nasu}},\ }\bibfield  {title} {\bibinfo {title} {{Majorana quasiparticles
  emergent in Kitaev spin liquid}},\ }\href
  {https://doi.org/10.1093/ptep/ptad115} {\bibfield  {journal} {\bibinfo
  {journal} {Prog. Theor. Exp. Phys.}\ ,\ \bibinfo {pages} {ptad115}} (\bibinfo
  {year} {2023})}\BibitemShut {NoStop}%
\bibitem [{\citenamefont {Sears}\ \emph {et~al.}(2020)\citenamefont {Sears},
  \citenamefont {Chern}, \citenamefont {Kim}, \citenamefont {Bereciartua},
  \citenamefont {Francoual}, \citenamefont {Kim},\ and\ \citenamefont
  {Kim}}]{Sears2020}%
  \BibitemOpen
  \bibfield  {author} {\bibinfo {author} {\bibfnamefont {J.~A.}\ \bibnamefont
  {Sears}}, \bibinfo {author} {\bibfnamefont {L.~E.}\ \bibnamefont {Chern}},
  \bibinfo {author} {\bibfnamefont {S.}~\bibnamefont {Kim}}, \bibinfo {author}
  {\bibfnamefont {P.~J.}\ \bibnamefont {Bereciartua}}, \bibinfo {author}
  {\bibfnamefont {S.}~\bibnamefont {Francoual}}, \bibinfo {author}
  {\bibfnamefont {Y.~B.}\ \bibnamefont {Kim}},\ and\ \bibinfo {author}
  {\bibfnamefont {Y.-J.}\ \bibnamefont {Kim}},\ }\bibfield  {title} {\bibinfo
  {title} {Ferromagnetic {K}itaev interaction and the origin of large magnetic
  anisotropy in {$\alpha$-\text{RuCl$_{3}$}}},\ }\href
  {https://doi.org/10.1038/s41567-020-0874-0} {\bibfield  {journal} {\bibinfo
  {journal} {Nat. Phys.}\ }\textbf {\bibinfo {volume} {16}},\ \bibinfo {pages}
  {837} (\bibinfo {year} {2020})}\BibitemShut {NoStop}%
\bibitem [{\citenamefont {Tennant}\ \emph {et~al.}(1995)\citenamefont
  {Tennant}, \citenamefont {Cowley}, \citenamefont {Nagler},\ and\
  \citenamefont {Tsvelik}}]{PhysRevB.52.13368}%
  \BibitemOpen
  \bibfield  {author} {\bibinfo {author} {\bibfnamefont {D.~A.}\ \bibnamefont
  {Tennant}}, \bibinfo {author} {\bibfnamefont {R.~A.}\ \bibnamefont {Cowley}},
  \bibinfo {author} {\bibfnamefont {S.~E.}\ \bibnamefont {Nagler}},\ and\
  \bibinfo {author} {\bibfnamefont {A.~M.}\ \bibnamefont {Tsvelik}},\
  }\bibfield  {title} {\bibinfo {title} {Measurement of the spin-excitation
  continuum in one-dimensional \text{K}\text{Cu}\text{F$_{3}$} using neutron
  scattering},\ }\href {https://doi.org/10.1103/PhysRevB.52.13368} {\bibfield
  {journal} {\bibinfo  {journal} {Phys. Rev. B}\ }\textbf {\bibinfo {volume}
  {52}},\ \bibinfo {pages} {13368} (\bibinfo {year} {1995})}\BibitemShut
  {NoStop}%
\bibitem [{\citenamefont {Lake}\ \emph {et~al.}(2013)\citenamefont {Lake},
  \citenamefont {Tennant}, \citenamefont {Caux}, \citenamefont {Barthel},
  \citenamefont {Schollw\"ock}, \citenamefont {Nagler},\ and\ \citenamefont
  {Frost}}]{PhysRevLett.111.137205}%
  \BibitemOpen
  \bibfield  {author} {\bibinfo {author} {\bibfnamefont {B.}~\bibnamefont
  {Lake}}, \bibinfo {author} {\bibfnamefont {D.~A.}\ \bibnamefont {Tennant}},
  \bibinfo {author} {\bibfnamefont {J.-S.}\ \bibnamefont {Caux}}, \bibinfo
  {author} {\bibfnamefont {T.}~\bibnamefont {Barthel}}, \bibinfo {author}
  {\bibfnamefont {U.}~\bibnamefont {Schollw\"ock}}, \bibinfo {author}
  {\bibfnamefont {S.~E.}\ \bibnamefont {Nagler}},\ and\ \bibinfo {author}
  {\bibfnamefont {C.~D.}\ \bibnamefont {Frost}},\ }\bibfield  {title} {\bibinfo
  {title} {Multispinon continua at zero and finite temperature in a near-ideal
  \text{H}eisenberg chain},\ }\href
  {https://doi.org/10.1103/PhysRevLett.111.137205} {\bibfield  {journal}
  {\bibinfo  {journal} {Phys. Rev. Lett.}\ }\textbf {\bibinfo {volume} {111}},\
  \bibinfo {pages} {137205} (\bibinfo {year} {2013})}\BibitemShut {NoStop}%
\bibitem [{\citenamefont {Anderson}(1973)}]{ANDERSON1973153}%
  \BibitemOpen
  \bibfield  {author} {\bibinfo {author} {\bibfnamefont {P.~W.}\ \bibnamefont
  {Anderson}},\ }\bibfield  {title} {\bibinfo {title} {Resonating valence
  bonds: A new kind of insulator?},\ }\href
  {https://doi.org/https://doi.org/10.1016/0025-5408(73)90167-0} {\bibfield
  {journal} {\bibinfo  {journal} {Mater. Res. Bull.}\ }\textbf {\bibinfo
  {volume} {8}},\ \bibinfo {pages} {153 } (\bibinfo {year} {1973})}\BibitemShut
  {NoStop}%
\bibitem [{\citenamefont {Kitaev}(2006)}]{Kitaev_2006}%
  \BibitemOpen
  \bibfield  {author} {\bibinfo {author} {\bibfnamefont {A.}~\bibnamefont
  {Kitaev}},\ }\bibfield  {title} {\bibinfo {title} {Anyons in an exactly
  solved model and beyond},\ }\href {https://doi.org/10.1016/j.aop.2005.10.005}
  {\bibfield  {journal} {\bibinfo  {journal} {Ann. Phys.}\ }\textbf {\bibinfo
  {volume} {321}},\ \bibinfo {pages} {2–111} (\bibinfo {year}
  {2006})}\BibitemShut {NoStop}%
\bibitem [{\citenamefont {Jeon}\ \emph {et~al.}(2024)\citenamefont {Jeon},
  \citenamefont {Wulferding}, \citenamefont {Choi}, \citenamefont {Lee},
  \citenamefont {Nam}, \citenamefont {Kim}, \citenamefont {Lee}, \citenamefont
  {Jang}, \citenamefont {Park}, \citenamefont {Lee}, \citenamefont {Choi},
  \citenamefont {Lee}, \citenamefont {Nojiri},\ and\ \citenamefont
  {Choi}}]{Jeon2024}%
  \BibitemOpen
  \bibfield  {author} {\bibinfo {author} {\bibfnamefont {S.}~\bibnamefont
  {Jeon}}, \bibinfo {author} {\bibfnamefont {D.}~\bibnamefont {Wulferding}},
  \bibinfo {author} {\bibfnamefont {Y.}~\bibnamefont {Choi}}, \bibinfo {author}
  {\bibfnamefont {S.}~\bibnamefont {Lee}}, \bibinfo {author} {\bibfnamefont
  {K.}~\bibnamefont {Nam}}, \bibinfo {author} {\bibfnamefont {K.~H.}\
  \bibnamefont {Kim}}, \bibinfo {author} {\bibfnamefont {M.}~\bibnamefont
  {Lee}}, \bibinfo {author} {\bibfnamefont {T.-H.}\ \bibnamefont {Jang}},
  \bibinfo {author} {\bibfnamefont {J.-H.}\ \bibnamefont {Park}}, \bibinfo
  {author} {\bibfnamefont {S.}~\bibnamefont {Lee}}, \bibinfo {author}
  {\bibfnamefont {S.}~\bibnamefont {Choi}}, \bibinfo {author} {\bibfnamefont
  {C.}~\bibnamefont {Lee}}, \bibinfo {author} {\bibfnamefont {H.}~\bibnamefont
  {Nojiri}},\ and\ \bibinfo {author} {\bibfnamefont {K.-Y.}\ \bibnamefont
  {Choi}},\ }\bibfield  {title} {\bibinfo {title} {One-ninth magnetization
  plateau stabilized by spin entanglement in a kagome antiferromagnet},\ }\href
  {https://doi.org/10.1038/s41567-023-02318-7} {\bibfield  {journal} {\bibinfo
  {journal} {Nat. Phys.}\ }\textbf {\bibinfo {volume} {20}},\ \bibinfo {pages}
  {435} (\bibinfo {year} {2024})}\BibitemShut {NoStop}%
\bibitem [{\citenamefont {Khuntia}\ \emph {et~al.}(2020)\citenamefont
  {Khuntia}, \citenamefont {Velazquez}, \citenamefont {Barth{\'e}lemy},
  \citenamefont {Bert}, \citenamefont {Kermarrec}, \citenamefont {Legros},
  \citenamefont {Bernu}, \citenamefont {Messio}, \citenamefont {Zorko},\ and\
  \citenamefont {Mendels}}]{Khuntia2020}%
  \BibitemOpen
  \bibfield  {author} {\bibinfo {author} {\bibfnamefont {P.}~\bibnamefont
  {Khuntia}}, \bibinfo {author} {\bibfnamefont {M.}~\bibnamefont {Velazquez}},
  \bibinfo {author} {\bibfnamefont {Q.}~\bibnamefont {Barth{\'e}lemy}},
  \bibinfo {author} {\bibfnamefont {F.}~\bibnamefont {Bert}}, \bibinfo {author}
  {\bibfnamefont {E.}~\bibnamefont {Kermarrec}}, \bibinfo {author}
  {\bibfnamefont {A.}~\bibnamefont {Legros}}, \bibinfo {author} {\bibfnamefont
  {B.}~\bibnamefont {Bernu}}, \bibinfo {author} {\bibfnamefont
  {L.}~\bibnamefont {Messio}}, \bibinfo {author} {\bibfnamefont
  {A.}~\bibnamefont {Zorko}},\ and\ \bibinfo {author} {\bibfnamefont
  {P.}~\bibnamefont {Mendels}},\ }\bibfield  {title} {\bibinfo {title} {Gapless
  ground state in the archetypal quantum kagome antiferromagnet
  {Zn\text{Cu$_{3}$}\text{(OH)$_{6}$}\text{Cl$_{2}$}}},\ }\href
  {https://doi.org/10.1038/s41567-020-0792-1} {\bibfield  {journal} {\bibinfo
  {journal} {Nat. Phys.}\ }\textbf {\bibinfo {volume} {16}},\ \bibinfo {pages}
  {469} (\bibinfo {year} {2020})}\BibitemShut {NoStop}%
\bibitem [{\citenamefont {Bag}\ \emph {et~al.}(2024)\citenamefont {Bag},
  \citenamefont {Xu}, \citenamefont {Sherman}, \citenamefont {Yadav},
  \citenamefont {Kolesnikov}, \citenamefont {Podlesnyak}, \citenamefont {Choi},
  \citenamefont {da~Silva}, \citenamefont {Moore},\ and\ \citenamefont
  {Haravifard}}]{PhysRevLett.133.266703}%
  \BibitemOpen
  \bibfield  {author} {\bibinfo {author} {\bibfnamefont {R.}~\bibnamefont
  {Bag}}, \bibinfo {author} {\bibfnamefont {S.}~\bibnamefont {Xu}}, \bibinfo
  {author} {\bibfnamefont {N.~E.}\ \bibnamefont {Sherman}}, \bibinfo {author}
  {\bibfnamefont {L.}~\bibnamefont {Yadav}}, \bibinfo {author} {\bibfnamefont
  {A.~I.}\ \bibnamefont {Kolesnikov}}, \bibinfo {author} {\bibfnamefont
  {A.~A.}\ \bibnamefont {Podlesnyak}}, \bibinfo {author} {\bibfnamefont
  {E.~S.}\ \bibnamefont {Choi}}, \bibinfo {author} {\bibfnamefont
  {I.}~\bibnamefont {da~Silva}}, \bibinfo {author} {\bibfnamefont {J.~E.}\
  \bibnamefont {Moore}},\ and\ \bibinfo {author} {\bibfnamefont
  {S.}~\bibnamefont {Haravifard}},\ }\bibfield  {title} {\bibinfo {title}
  {Evidence of dirac quantum spin liquid in
  \text{Yb}\text{Zn$_{2}$}\text{Ga}\text{O$_{5}$}},\ }\href
  {https://doi.org/10.1103/PhysRevLett.133.266703} {\bibfield  {journal}
  {\bibinfo  {journal} {Phys. Rev. Lett.}\ }\textbf {\bibinfo {volume} {133}},\
  \bibinfo {pages} {266703} (\bibinfo {year} {2024})}\BibitemShut {NoStop}%
\bibitem [{\citenamefont {Li}(2019)}]{qute}%
  \BibitemOpen
  \bibfield  {author} {\bibinfo {author} {\bibfnamefont {Y.}~\bibnamefont
  {Li}},\ }\bibfield  {title} {\bibinfo {title}
  {Yb\text{M}g\text{G}a\text{O$_{4}$}: A triangular-lattice quantum spin liquid
  candidate},\ }\href {https://doi.org/https://doi.org/10.1002/qute.201900089}
  {\bibfield  {journal} {\bibinfo  {journal} {Adv. Quantum Technol.}\ }\textbf
  {\bibinfo {volume} {2}},\ \bibinfo {pages} {1900089} (\bibinfo {year}
  {2019})}\BibitemShut {NoStop}%
\bibitem [{\citenamefont {Arh}\ \emph {et~al.}(2022)\citenamefont {Arh},
  \citenamefont {Sana}, \citenamefont {Pregelj}, \citenamefont {Khuntia},
  \citenamefont {Jagli{\v{c}}i{\'{c}}}, \citenamefont {Le}, \citenamefont
  {Biswas}, \citenamefont {Manuel}, \citenamefont {Mangin-Thro}, \citenamefont
  {Ozarowski},\ and\ \citenamefont {Zorko}}]{Arh2022}%
  \BibitemOpen
  \bibfield  {author} {\bibinfo {author} {\bibfnamefont {T.}~\bibnamefont
  {Arh}}, \bibinfo {author} {\bibfnamefont {B.}~\bibnamefont {Sana}}, \bibinfo
  {author} {\bibfnamefont {M.}~\bibnamefont {Pregelj}}, \bibinfo {author}
  {\bibfnamefont {P.}~\bibnamefont {Khuntia}}, \bibinfo {author} {\bibfnamefont
  {Z.}~\bibnamefont {Jagli{\v{c}}i{\'{c}}}}, \bibinfo {author} {\bibfnamefont
  {M.~D.}\ \bibnamefont {Le}}, \bibinfo {author} {\bibfnamefont {P.~K.}\
  \bibnamefont {Biswas}}, \bibinfo {author} {\bibfnamefont {P.}~\bibnamefont
  {Manuel}}, \bibinfo {author} {\bibfnamefont {L.}~\bibnamefont {Mangin-Thro}},
  \bibinfo {author} {\bibfnamefont {A.}~\bibnamefont {Ozarowski}},\ and\
  \bibinfo {author} {\bibfnamefont {A.}~\bibnamefont {Zorko}},\ }\bibfield
  {title} {\bibinfo {title} {The \text{I}sing triangular-lattice
  antiferromagnet neodymium heptatantalate as a quantum spin liquid
  candidate},\ }\href {https://doi.org/10.1038/s41563-021-01169-y} {\bibfield
  {journal} {\bibinfo  {journal} {Nat. Mater.}\ }\textbf {\bibinfo {volume}
  {21}},\ \bibinfo {pages} {416} (\bibinfo {year} {2022})}\BibitemShut
  {NoStop}%
\bibitem [{\citenamefont {Bordelon}\ \emph {et~al.}(2019)\citenamefont
  {Bordelon}, \citenamefont {Kenney}, \citenamefont {Liu}, \citenamefont
  {Hogan}, \citenamefont {Posthuma}, \citenamefont {Kavand}, \citenamefont
  {Lyu}, \citenamefont {Sherwin}, \citenamefont {Butch}, \citenamefont {Brown},
  \citenamefont {Graf}, \citenamefont {Balents},\ and\ \citenamefont
  {Wilson}}]{Bordelon2019}%
  \BibitemOpen
  \bibfield  {author} {\bibinfo {author} {\bibfnamefont {M.~M.}\ \bibnamefont
  {Bordelon}}, \bibinfo {author} {\bibfnamefont {E.}~\bibnamefont {Kenney}},
  \bibinfo {author} {\bibfnamefont {C.}~\bibnamefont {Liu}}, \bibinfo {author}
  {\bibfnamefont {T.}~\bibnamefont {Hogan}}, \bibinfo {author} {\bibfnamefont
  {L.}~\bibnamefont {Posthuma}}, \bibinfo {author} {\bibfnamefont
  {M.}~\bibnamefont {Kavand}}, \bibinfo {author} {\bibfnamefont
  {Y.}~\bibnamefont {Lyu}}, \bibinfo {author} {\bibfnamefont {M.}~\bibnamefont
  {Sherwin}}, \bibinfo {author} {\bibfnamefont {N.~P.}\ \bibnamefont {Butch}},
  \bibinfo {author} {\bibfnamefont {C.}~\bibnamefont {Brown}}, \bibinfo
  {author} {\bibfnamefont {M.~J.}\ \bibnamefont {Graf}}, \bibinfo {author}
  {\bibfnamefont {L.}~\bibnamefont {Balents}},\ and\ \bibinfo {author}
  {\bibfnamefont {S.~D.}\ \bibnamefont {Wilson}},\ }\bibfield  {title}
  {\bibinfo {title} {Field-tunable quantum disordered ground state in the
  triangular-lattice antiferromagnet {NaYbO$_{2}$}},\ }\href
  {https://doi.org/10.1038/s41567-019-0594-5} {\bibfield  {journal} {\bibinfo
  {journal} {Nat. Phys.}\ }\textbf {\bibinfo {volume} {15}},\ \bibinfo {pages}
  {1058} (\bibinfo {year} {2019})}\BibitemShut {NoStop}%
\bibitem [{\citenamefont {Chillal}\ \emph {et~al.}(2020)\citenamefont
  {Chillal}, \citenamefont {Iqbal}, \citenamefont {Jeschke}, \citenamefont
  {Rodriguez-Rivera}, \citenamefont {Bewley}, \citenamefont {Manuel},
  \citenamefont {Khalyavin}, \citenamefont {Steffens}, \citenamefont {Thomale},
  \citenamefont {Islam}, \citenamefont {Reuther},\ and\ \citenamefont
  {Lake}}]{Chillal2020}%
  \BibitemOpen
  \bibfield  {author} {\bibinfo {author} {\bibfnamefont {S.}~\bibnamefont
  {Chillal}}, \bibinfo {author} {\bibfnamefont {Y.}~\bibnamefont {Iqbal}},
  \bibinfo {author} {\bibfnamefont {H.~O.}\ \bibnamefont {Jeschke}}, \bibinfo
  {author} {\bibfnamefont {J.~A.}\ \bibnamefont {Rodriguez-Rivera}}, \bibinfo
  {author} {\bibfnamefont {R.}~\bibnamefont {Bewley}}, \bibinfo {author}
  {\bibfnamefont {P.}~\bibnamefont {Manuel}}, \bibinfo {author} {\bibfnamefont
  {D.}~\bibnamefont {Khalyavin}}, \bibinfo {author} {\bibfnamefont
  {P.}~\bibnamefont {Steffens}}, \bibinfo {author} {\bibfnamefont
  {R.}~\bibnamefont {Thomale}}, \bibinfo {author} {\bibfnamefont {A.~T. M.~N.}\
  \bibnamefont {Islam}}, \bibinfo {author} {\bibfnamefont {J.}~\bibnamefont
  {Reuther}},\ and\ \bibinfo {author} {\bibfnamefont {B.}~\bibnamefont
  {Lake}},\ }\bibfield  {title} {\bibinfo {title} {Evidence for a
  three-dimensional quantum spin liquid in
  \text{Pb}\text{Cu}\text{Te$_{2}$}\text{O$_{6}$}},\ }\href
  {https://doi.org/10.1038/s41467-020-15594-1} {\bibfield  {journal} {\bibinfo
  {journal} {Nat. Commun.}\ }\textbf {\bibinfo {volume} {11}},\ \bibinfo
  {pages} {2348} (\bibinfo {year} {2020})}\BibitemShut {NoStop}%
\bibitem [{\citenamefont {Khuntia}\ \emph {et~al.}(2016)\citenamefont
  {Khuntia}, \citenamefont {Bert}, \citenamefont {Mendels}, \citenamefont
  {Koteswararao}, \citenamefont {Mahajan}, \citenamefont {Baenitz},
  \citenamefont {Chou}, \citenamefont {Baines}, \citenamefont {Amato},\ and\
  \citenamefont {Furukawa}}]{PhysRevLett.116.107203}%
  \BibitemOpen
  \bibfield  {author} {\bibinfo {author} {\bibfnamefont {P.}~\bibnamefont
  {Khuntia}}, \bibinfo {author} {\bibfnamefont {F.}~\bibnamefont {Bert}},
  \bibinfo {author} {\bibfnamefont {P.}~\bibnamefont {Mendels}}, \bibinfo
  {author} {\bibfnamefont {B.}~\bibnamefont {Koteswararao}}, \bibinfo {author}
  {\bibfnamefont {A.~V.}\ \bibnamefont {Mahajan}}, \bibinfo {author}
  {\bibfnamefont {M.}~\bibnamefont {Baenitz}}, \bibinfo {author} {\bibfnamefont
  {F.~C.}\ \bibnamefont {Chou}}, \bibinfo {author} {\bibfnamefont
  {C.}~\bibnamefont {Baines}}, \bibinfo {author} {\bibfnamefont
  {A.}~\bibnamefont {Amato}},\ and\ \bibinfo {author} {\bibfnamefont
  {Y.}~\bibnamefont {Furukawa}},\ }\bibfield  {title} {\bibinfo {title} {Spin
  liquid state in the 3\text{D} frustrated antiferromagnet
  {Pb}{Cu}\text{Te$_{2}$}\text{O$_{6}$}: {NMR} and {M}uon spin relaxation
  studies},\ }\href {https://doi.org/10.1103/PhysRevLett.116.107203} {\bibfield
   {journal} {\bibinfo  {journal} {Phys. Rev. Lett.}\ }\textbf {\bibinfo
  {volume} {116}},\ \bibinfo {pages} {107203} (\bibinfo {year}
  {2016})}\BibitemShut {NoStop}%
\bibitem [{\citenamefont {Plumb}\ \emph {et~al.}(2019)\citenamefont {Plumb},
  \citenamefont {Changlani}, \citenamefont {Scheie}, \citenamefont {Zhang},
  \citenamefont {Krizan}, \citenamefont {Rodriguez-Rivera}, \citenamefont
  {Qiu}, \citenamefont {Winn}, \citenamefont {Cava},\ and\ \citenamefont
  {Broholm}}]{Plumb2019}%
  \BibitemOpen
  \bibfield  {author} {\bibinfo {author} {\bibfnamefont {K.~W.}\ \bibnamefont
  {Plumb}}, \bibinfo {author} {\bibfnamefont {H.~J.}\ \bibnamefont
  {Changlani}}, \bibinfo {author} {\bibfnamefont {A.}~\bibnamefont {Scheie}},
  \bibinfo {author} {\bibfnamefont {S.}~\bibnamefont {Zhang}}, \bibinfo
  {author} {\bibfnamefont {J.~W.}\ \bibnamefont {Krizan}}, \bibinfo {author}
  {\bibfnamefont {J.~A.}\ \bibnamefont {Rodriguez-Rivera}}, \bibinfo {author}
  {\bibfnamefont {Y.}~\bibnamefont {Qiu}}, \bibinfo {author} {\bibfnamefont
  {B.}~\bibnamefont {Winn}}, \bibinfo {author} {\bibfnamefont {R.~J.}\
  \bibnamefont {Cava}},\ and\ \bibinfo {author} {\bibfnamefont {C.~L.}\
  \bibnamefont {Broholm}},\ }\bibfield  {title} {\bibinfo {title} {Continuum of
  quantum fluctuations in a three-dimensional s{\thinspace}={\thinspace}1
  {H}eisenberg magnet},\ }\href {https://doi.org/10.1038/s41567-018-0317-3}
  {\bibfield  {journal} {\bibinfo  {journal} {Nat. Phys.}\ }\textbf {\bibinfo
  {volume} {15}},\ \bibinfo {pages} {54} (\bibinfo {year} {2019})}\BibitemShut
  {NoStop}%
\bibitem [{\citenamefont {Gao}\ \emph {et~al.}(2019)\citenamefont {Gao},
  \citenamefont {Chen}, \citenamefont {Tam}, \citenamefont {Huang},
  \citenamefont {Sasmal}, \citenamefont {Adroja}, \citenamefont {Ye},
  \citenamefont {Cao}, \citenamefont {Sala}, \citenamefont {Stone},
  \citenamefont {Baines}, \citenamefont {Verezhak}, \citenamefont {Hu},
  \citenamefont {Chung}, \citenamefont {Xu}, \citenamefont {Cheong},
  \citenamefont {Nallaiyan}, \citenamefont {Spagna}, \citenamefont {Maple},
  \citenamefont {Nevidomskyy}, \citenamefont {Morosan}, \citenamefont {Chen},\
  and\ \citenamefont {Dai}}]{Gao2019}%
  \BibitemOpen
  \bibfield  {author} {\bibinfo {author} {\bibfnamefont {B.}~\bibnamefont
  {Gao}}, \bibinfo {author} {\bibfnamefont {T.}~\bibnamefont {Chen}}, \bibinfo
  {author} {\bibfnamefont {D.~W.}\ \bibnamefont {Tam}}, \bibinfo {author}
  {\bibfnamefont {C.-L.}\ \bibnamefont {Huang}}, \bibinfo {author}
  {\bibfnamefont {K.}~\bibnamefont {Sasmal}}, \bibinfo {author} {\bibfnamefont
  {D.~T.}\ \bibnamefont {Adroja}}, \bibinfo {author} {\bibfnamefont
  {F.}~\bibnamefont {Ye}}, \bibinfo {author} {\bibfnamefont {H.}~\bibnamefont
  {Cao}}, \bibinfo {author} {\bibfnamefont {G.}~\bibnamefont {Sala}}, \bibinfo
  {author} {\bibfnamefont {M.~B.}\ \bibnamefont {Stone}}, \bibinfo {author}
  {\bibfnamefont {C.}~\bibnamefont {Baines}}, \bibinfo {author} {\bibfnamefont
  {J.~A.~T.}\ \bibnamefont {Verezhak}}, \bibinfo {author} {\bibfnamefont
  {H.}~\bibnamefont {Hu}}, \bibinfo {author} {\bibfnamefont {J.-H.}\
  \bibnamefont {Chung}}, \bibinfo {author} {\bibfnamefont {X.}~\bibnamefont
  {Xu}}, \bibinfo {author} {\bibfnamefont {S.-W.}\ \bibnamefont {Cheong}},
  \bibinfo {author} {\bibfnamefont {M.}~\bibnamefont {Nallaiyan}}, \bibinfo
  {author} {\bibfnamefont {S.}~\bibnamefont {Spagna}}, \bibinfo {author}
  {\bibfnamefont {M.~B.}\ \bibnamefont {Maple}}, \bibinfo {author}
  {\bibfnamefont {A.~H.}\ \bibnamefont {Nevidomskyy}}, \bibinfo {author}
  {\bibfnamefont {E.}~\bibnamefont {Morosan}}, \bibinfo {author} {\bibfnamefont
  {G.}~\bibnamefont {Chen}},\ and\ \bibinfo {author} {\bibfnamefont
  {P.}~\bibnamefont {Dai}},\ }\bibfield  {title} {\bibinfo {title}
  {Experimental signatures of a three-dimensional quantum spin liquid in
  effective spin-1/2 \text{Ce$_{2}$}\text{Zr$_{2}$}\text{O$_{7}$} pyrochlore},\
  }\href {https://doi.org/10.1038/s41567-019-0577-6} {\bibfield  {journal}
  {\bibinfo  {journal} {Nat. Phys.}\ }\textbf {\bibinfo {volume} {15}},\
  \bibinfo {pages} {1052} (\bibinfo {year} {2019})}\BibitemShut {NoStop}%
\bibitem [{\citenamefont {Wen}(2002)}]{PhysRevB.65.165113}%
  \BibitemOpen
  \bibfield  {author} {\bibinfo {author} {\bibfnamefont {X.-G.}\ \bibnamefont
  {Wen}},\ }\bibfield  {title} {\bibinfo {title} {Quantum orders and symmetric
  spin liquids},\ }\href {https://doi.org/10.1103/PhysRevB.65.165113}
  {\bibfield  {journal} {\bibinfo  {journal} {Phys. Rev. B}\ }\textbf {\bibinfo
  {volume} {65}},\ \bibinfo {pages} {165113} (\bibinfo {year}
  {2002})}\BibitemShut {NoStop}%
\bibitem [{\citenamefont {Kalmeyer}\ and\ \citenamefont
  {Laughlin}(1987)}]{PhysRevLett.59.2095}%
  \BibitemOpen
  \bibfield  {author} {\bibinfo {author} {\bibfnamefont {V.}~\bibnamefont
  {Kalmeyer}}\ and\ \bibinfo {author} {\bibfnamefont {R.~B.}\ \bibnamefont
  {Laughlin}},\ }\bibfield  {title} {\bibinfo {title} {Equivalence of the
  resonating-valence-bond and fractional quantum \text{H}all states},\ }\href
  {https://doi.org/10.1103/PhysRevLett.59.2095} {\bibfield  {journal} {\bibinfo
   {journal} {Phys. Rev. Lett.}\ }\textbf {\bibinfo {volume} {59}},\ \bibinfo
  {pages} {2095} (\bibinfo {year} {1987})}\BibitemShut {NoStop}%
\bibitem [{\citenamefont {Kadow}\ \emph {et~al.}(2022)\citenamefont {Kadow},
  \citenamefont {Vanderstraeten},\ and\ \citenamefont
  {Knap}}]{PhysRevB.106.094417}%
  \BibitemOpen
  \bibfield  {author} {\bibinfo {author} {\bibfnamefont {W.}~\bibnamefont
  {Kadow}}, \bibinfo {author} {\bibfnamefont {L.}~\bibnamefont
  {Vanderstraeten}},\ and\ \bibinfo {author} {\bibfnamefont {M.}~\bibnamefont
  {Knap}},\ }\bibfield  {title} {\bibinfo {title} {Hole spectral function of a
  chiral spin liquid in the triangular lattice hubbard model},\ }\href
  {https://doi.org/10.1103/PhysRevB.106.094417} {\bibfield  {journal} {\bibinfo
   {journal} {Phys. Rev. B}\ }\textbf {\bibinfo {volume} {106}},\ \bibinfo
  {pages} {094417} (\bibinfo {year} {2022})}\BibitemShut {NoStop}%
\bibitem [{\citenamefont {Zhang}\ \emph {et~al.}(2024)\citenamefont {Zhang},
  \citenamefont {Huang}, \citenamefont {Wu}, \citenamefont {Sheng},\ and\
  \citenamefont {Gong}}]{PhysRevB.109.125146}%
  \BibitemOpen
  \bibfield  {author} {\bibinfo {author} {\bibfnamefont {X.-T.}\ \bibnamefont
  {Zhang}}, \bibinfo {author} {\bibfnamefont {Y.}~\bibnamefont {Huang}},
  \bibinfo {author} {\bibfnamefont {H.-Q.}\ \bibnamefont {Wu}}, \bibinfo
  {author} {\bibfnamefont {D.~N.}\ \bibnamefont {Sheng}},\ and\ \bibinfo
  {author} {\bibfnamefont {S.-S.}\ \bibnamefont {Gong}},\ }\bibfield  {title}
  {\bibinfo {title} {Chiral spin liquid and quantum phase diagram of
  spin-$\frac{1}{2}
  {J}_{1}\text{\ensuremath{-}}{J}_{2}\text{\ensuremath{-}}{J}_{\ensuremath{\chi}}$
  model on the square lattice},\ }\href
  {https://doi.org/10.1103/PhysRevB.109.125146} {\bibfield  {journal} {\bibinfo
   {journal} {Phys. Rev. B}\ }\textbf {\bibinfo {volume} {109}},\ \bibinfo
  {pages} {125146} (\bibinfo {year} {2024})}\BibitemShut {NoStop}%
\bibitem [{\citenamefont {Zhu}(2024)}]{PhysRevB.110.L041113}%
  \BibitemOpen
  \bibfield  {author} {\bibinfo {author} {\bibfnamefont {Z.}~\bibnamefont
  {Zhu}},\ }\bibfield  {title} {\bibinfo {title} {Chiral spin liquid versus
  mott antiferromagnetism in the triangular-lattice hubbard model},\ }\href
  {https://doi.org/10.1103/PhysRevB.110.L041113} {\bibfield  {journal}
  {\bibinfo  {journal} {Phys. Rev. B}\ }\textbf {\bibinfo {volume} {110}},\
  \bibinfo {pages} {L041113} (\bibinfo {year} {2024})}\BibitemShut {NoStop}%
\bibitem [{\citenamefont {Bauer}\ \emph {et~al.}(2014)\citenamefont {Bauer},
  \citenamefont {Cincio}, \citenamefont {Keller}, \citenamefont {Dolfi},
  \citenamefont {Vidal}, \citenamefont {Trebst},\ and\ \citenamefont
  {Ludwig}}]{Bauer2014}%
  \BibitemOpen
  \bibfield  {author} {\bibinfo {author} {\bibfnamefont {B.}~\bibnamefont
  {Bauer}}, \bibinfo {author} {\bibfnamefont {L.}~\bibnamefont {Cincio}},
  \bibinfo {author} {\bibfnamefont {B.~P.}\ \bibnamefont {Keller}}, \bibinfo
  {author} {\bibfnamefont {M.}~\bibnamefont {Dolfi}}, \bibinfo {author}
  {\bibfnamefont {G.}~\bibnamefont {Vidal}}, \bibinfo {author} {\bibfnamefont
  {S.}~\bibnamefont {Trebst}},\ and\ \bibinfo {author} {\bibfnamefont
  {A.~W.~W.}\ \bibnamefont {Ludwig}},\ }\bibfield  {title} {\bibinfo {title}
  {Chiral spin liquid and emergent anyons in a kagome lattice mott insulator},\
  }\href {https://doi.org/10.1038/ncomms6137} {\bibfield  {journal} {\bibinfo
  {journal} {Nat. Commun.}\ }\textbf {\bibinfo {volume} {5}},\ \bibinfo {pages}
  {5137} (\bibinfo {year} {2014})}\BibitemShut {NoStop}%
\bibitem [{\citenamefont {Huang}\ \emph {et~al.}(2021)\citenamefont {Huang},
  \citenamefont {Dong}, \citenamefont {Sheng},\ and\ \citenamefont
  {Ting}}]{PhysRevB.103.L041108}%
  \BibitemOpen
  \bibfield  {author} {\bibinfo {author} {\bibfnamefont {Y.}~\bibnamefont
  {Huang}}, \bibinfo {author} {\bibfnamefont {X.-Y.}\ \bibnamefont {Dong}},
  \bibinfo {author} {\bibfnamefont {D.~N.}\ \bibnamefont {Sheng}},\ and\
  \bibinfo {author} {\bibfnamefont {C.~S.}\ \bibnamefont {Ting}},\ }\bibfield
  {title} {\bibinfo {title} {Quantum phase diagram and chiral spin liquid in
  the extended spin-$\frac{1}{2}$ honeycomb xy model},\ }\href
  {https://doi.org/10.1103/PhysRevB.103.L041108} {\bibfield  {journal}
  {\bibinfo  {journal} {Phys. Rev. B}\ }\textbf {\bibinfo {volume} {103}},\
  \bibinfo {pages} {L041108} (\bibinfo {year} {2021})}\BibitemShut {NoStop}%
\bibitem [{\citenamefont {Lozano-G{\'o}mez}\ \emph {et~al.}(2024)\citenamefont
  {Lozano-G{\'o}mez}, \citenamefont {Iqbal},\ and\ \citenamefont
  {Vojta}}]{Lozano2024}%
  \BibitemOpen
  \bibfield  {author} {\bibinfo {author} {\bibfnamefont {D.}~\bibnamefont
  {Lozano-G{\'o}mez}}, \bibinfo {author} {\bibfnamefont {Y.}~\bibnamefont
  {Iqbal}},\ and\ \bibinfo {author} {\bibfnamefont {M.}~\bibnamefont {Vojta}},\
  }\bibfield  {title} {\bibinfo {title} {A classical chiral spin liquid from
  chiral interactions on the pyrochlore lattice},\ }\href
  {https://doi.org/10.1038/s41467-024-54558-7} {\bibfield  {journal} {\bibinfo
  {journal} {Nat. Commun.}\ }\textbf {\bibinfo {volume} {15}},\ \bibinfo
  {pages} {10162} (\bibinfo {year} {2024})}\BibitemShut {NoStop}%
\bibitem [{\citenamefont {Fancelli}\ \emph {et~al.}(2025)\citenamefont
  {Fancelli}, \citenamefont {Flores-Calder\'on}, \citenamefont {Benton},
  \citenamefont {Lake}, \citenamefont {Moessner},\ and\ \citenamefont
  {Reuther}}]{PhysRevB.111.134413}%
  \BibitemOpen
  \bibfield  {author} {\bibinfo {author} {\bibfnamefont {A.}~\bibnamefont
  {Fancelli}}, \bibinfo {author} {\bibfnamefont {R.}~\bibnamefont
  {Flores-Calder\'on}}, \bibinfo {author} {\bibfnamefont {O.}~\bibnamefont
  {Benton}}, \bibinfo {author} {\bibfnamefont {B.}~\bibnamefont {Lake}},
  \bibinfo {author} {\bibfnamefont {R.}~\bibnamefont {Moessner}},\ and\
  \bibinfo {author} {\bibfnamefont {J.}~\bibnamefont {Reuther}},\ }\bibfield
  {title} {\bibinfo {title} {Fragile spin liquid in three dimensions},\ }\href
  {https://doi.org/10.1103/PhysRevB.111.134413} {\bibfield  {journal} {\bibinfo
   {journal} {Phys. Rev. B}\ }\textbf {\bibinfo {volume} {111}},\ \bibinfo
  {pages} {134413} (\bibinfo {year} {2025})}\BibitemShut {NoStop}%
\bibitem [{\citenamefont {Yan}\ \emph {et~al.}(2024)\citenamefont {Yan},
  \citenamefont {Benton}, \citenamefont {Moessner},\ and\ \citenamefont
  {Nevidomskyy}}]{PhysRevB.110.L020402}%
  \BibitemOpen
  \bibfield  {author} {\bibinfo {author} {\bibfnamefont {H.}~\bibnamefont
  {Yan}}, \bibinfo {author} {\bibfnamefont {O.}~\bibnamefont {Benton}},
  \bibinfo {author} {\bibfnamefont {R.}~\bibnamefont {Moessner}},\ and\
  \bibinfo {author} {\bibfnamefont {A.~H.}\ \bibnamefont {Nevidomskyy}},\
  }\bibfield  {title} {\bibinfo {title} {Classification of classical spin
  liquids: Typology and resulting landscape},\ }\href
  {https://doi.org/10.1103/PhysRevB.110.L020402} {\bibfield  {journal}
  {\bibinfo  {journal} {Phys. Rev. B}\ }\textbf {\bibinfo {volume} {110}},\
  \bibinfo {pages} {L020402} (\bibinfo {year} {2024})}\BibitemShut {NoStop}%
\bibitem [{\citenamefont {Niggemann}\ \emph {et~al.}(2019)\citenamefont
  {Niggemann}, \citenamefont {Hering},\ and\ \citenamefont
  {Reuther}}]{Niggemann_2020}%
  \BibitemOpen
  \bibfield  {author} {\bibinfo {author} {\bibfnamefont {N.}~\bibnamefont
  {Niggemann}}, \bibinfo {author} {\bibfnamefont {M.}~\bibnamefont {Hering}},\
  and\ \bibinfo {author} {\bibfnamefont {J.}~\bibnamefont {Reuther}},\
  }\bibfield  {title} {\bibinfo {title} {Classical spiral spin liquids as a
  possible route to quantum spin liquids},\ }\href
  {https://doi.org/10.1088/1361-648X/ab4480} {\bibfield  {journal} {\bibinfo
  {journal} {J. Condens. Matter Phys.}\ }\textbf {\bibinfo {volume} {32}},\
  \bibinfo {pages} {024001} (\bibinfo {year} {2019})}\BibitemShut {NoStop}%
\bibitem [{\citenamefont {Isakov}\ \emph {et~al.}(2004)\citenamefont {Isakov},
  \citenamefont {Gregor}, \citenamefont {Moessner},\ and\ \citenamefont
  {Sondhi}}]{PhysRevLett.93.167204}%
  \BibitemOpen
  \bibfield  {author} {\bibinfo {author} {\bibfnamefont {S.~V.}\ \bibnamefont
  {Isakov}}, \bibinfo {author} {\bibfnamefont {K.}~\bibnamefont {Gregor}},
  \bibinfo {author} {\bibfnamefont {R.}~\bibnamefont {Moessner}},\ and\
  \bibinfo {author} {\bibfnamefont {S.~L.}\ \bibnamefont {Sondhi}},\ }\bibfield
   {title} {\bibinfo {title} {Dipolar spin correlations in classical pyrochlore
  magnets},\ }\href {https://doi.org/10.1103/PhysRevLett.93.167204} {\bibfield
  {journal} {\bibinfo  {journal} {Phys. Rev. Lett.}\ }\textbf {\bibinfo
  {volume} {93}},\ \bibinfo {pages} {167204} (\bibinfo {year}
  {2004})}\BibitemShut {NoStop}%
\bibitem [{\citenamefont {Castelnovo}\ \emph {et~al.}(2012)\citenamefont
  {Castelnovo}, \citenamefont {Moessner},\ and\ \citenamefont
  {Sondhi}}]{annurev}%
  \BibitemOpen
  \bibfield  {author} {\bibinfo {author} {\bibfnamefont {C.}~\bibnamefont
  {Castelnovo}}, \bibinfo {author} {\bibfnamefont {R.}~\bibnamefont
  {Moessner}},\ and\ \bibinfo {author} {\bibfnamefont {S.}~\bibnamefont
  {Sondhi}},\ }\bibfield  {title} {\bibinfo {title} {Spin ice,
  fractionalization, and topological order},\ }\href
  {https://doi.org/https://doi.org/10.1146/annurev-conmatphys-020911-125058}
  {\bibfield  {journal} {\bibinfo  {journal} {Annu. Rev. Condens. Matter
  Phys.}\ }\textbf {\bibinfo {volume} {3}},\ \bibinfo {pages} {35} (\bibinfo
  {year} {2012})}\BibitemShut {NoStop}%
\bibitem [{\citenamefont {\ifmmode \check{Z}\else
  \v{Z}\fi{}ivkovi\ifmmode~\acute{c}\else \'{c}\fi{}}\ \emph
  {et~al.}(2021)\citenamefont {\ifmmode \check{Z}\else
  \v{Z}\fi{}ivkovi\ifmmode~\acute{c}\else \'{c}\fi{}}, \citenamefont {Favre},
  \citenamefont {Salazar~Mejia}, \citenamefont {Jeschke}, \citenamefont
  {Magrez}, \citenamefont {Dabholkar}, \citenamefont {Noculak}, \citenamefont
  {Freitas}, \citenamefont {Jeong}, \citenamefont {Hegde}, \citenamefont
  {Testa}, \citenamefont {Babkevich}, \citenamefont {Su}, \citenamefont
  {Manuel}, \citenamefont {Luetkens}, \citenamefont {Baines}, \citenamefont
  {Baker}, \citenamefont {Wosnitza}, \citenamefont {Zaharko}, \citenamefont
  {Iqbal}, \citenamefont {Reuther},\ and\ \citenamefont
  {R\o{}nnow}}]{PhysRevLett.127.157204}%
  \BibitemOpen
  \bibfield  {author} {\bibinfo {author} {\bibfnamefont {I.}~\bibnamefont
  {\ifmmode \check{Z}\else \v{Z}\fi{}ivkovi\ifmmode~\acute{c}\else
  \'{c}\fi{}}}, \bibinfo {author} {\bibfnamefont {V.}~\bibnamefont {Favre}},
  \bibinfo {author} {\bibfnamefont {C.}~\bibnamefont {Salazar~Mejia}}, \bibinfo
  {author} {\bibfnamefont {H.~O.}\ \bibnamefont {Jeschke}}, \bibinfo {author}
  {\bibfnamefont {A.}~\bibnamefont {Magrez}}, \bibinfo {author} {\bibfnamefont
  {B.}~\bibnamefont {Dabholkar}}, \bibinfo {author} {\bibfnamefont
  {V.}~\bibnamefont {Noculak}}, \bibinfo {author} {\bibfnamefont {R.~S.}\
  \bibnamefont {Freitas}}, \bibinfo {author} {\bibfnamefont {M.}~\bibnamefont
  {Jeong}}, \bibinfo {author} {\bibfnamefont {N.~G.}\ \bibnamefont {Hegde}},
  \bibinfo {author} {\bibfnamefont {L.}~\bibnamefont {Testa}}, \bibinfo
  {author} {\bibfnamefont {P.}~\bibnamefont {Babkevich}}, \bibinfo {author}
  {\bibfnamefont {Y.}~\bibnamefont {Su}}, \bibinfo {author} {\bibfnamefont
  {P.}~\bibnamefont {Manuel}}, \bibinfo {author} {\bibfnamefont
  {H.}~\bibnamefont {Luetkens}}, \bibinfo {author} {\bibfnamefont
  {C.}~\bibnamefont {Baines}}, \bibinfo {author} {\bibfnamefont {P.~J.}\
  \bibnamefont {Baker}}, \bibinfo {author} {\bibfnamefont {J.}~\bibnamefont
  {Wosnitza}}, \bibinfo {author} {\bibfnamefont {O.}~\bibnamefont {Zaharko}},
  \bibinfo {author} {\bibfnamefont {Y.}~\bibnamefont {Iqbal}}, \bibinfo
  {author} {\bibfnamefont {J.}~\bibnamefont {Reuther}},\ and\ \bibinfo {author}
  {\bibfnamefont {H.~M.}\ \bibnamefont {R\o{}nnow}},\ }\bibfield  {title}
  {\bibinfo {title} {Magnetic field induced quantum spin liquid in the two
  coupled trillium lattices of
  \text{K$_{2}$}\text{Ni$_{2}$}\text{(SO$_{4}$)$_{3}$}},\ }\href
  {https://doi.org/10.1103/PhysRevLett.127.157204} {\bibfield  {journal}
  {\bibinfo  {journal} {Phys. Rev. Lett.}\ }\textbf {\bibinfo {volume} {127}},\
  \bibinfo {pages} {157204} (\bibinfo {year} {2021})}\BibitemShut {NoStop}%
\bibitem [{\citenamefont {Hopkinson}\ and\ \citenamefont
  {Kee}(2006)}]{PhysRevB.74.224441}%
  \BibitemOpen
  \bibfield  {author} {\bibinfo {author} {\bibfnamefont {J.~M.}\ \bibnamefont
  {Hopkinson}}\ and\ \bibinfo {author} {\bibfnamefont {H.-Y.}\ \bibnamefont
  {Kee}},\ }\bibfield  {title} {\bibinfo {title} {Geometric frustration
  inherent to the trillium lattice, a sublattice of the {B}20 structure},\
  }\href {https://doi.org/10.1103/PhysRevB.74.224441} {\bibfield  {journal}
  {\bibinfo  {journal} {Phys. Rev. B}\ }\textbf {\bibinfo {volume} {74}},\
  \bibinfo {pages} {224441} (\bibinfo {year} {2006})}\BibitemShut {NoStop}%
\bibitem [{\citenamefont {Boya}\ \emph {et~al.}(2022)\citenamefont {Boya},
  \citenamefont {Nam}, \citenamefont {Kargeti}, \citenamefont {Jain},
  \citenamefont {Kumar}, \citenamefont {Panda}, \citenamefont {Yusuf},
  \citenamefont {Paulose}, \citenamefont {Voma}, \citenamefont {Kermarrec},
  \citenamefont {Kim},\ and\ \citenamefont {Koteswararao}}]{10.1063/5.0096942}%
  \BibitemOpen
  \bibfield  {author} {\bibinfo {author} {\bibfnamefont {K.}~\bibnamefont
  {Boya}}, \bibinfo {author} {\bibfnamefont {K.}~\bibnamefont {Nam}}, \bibinfo
  {author} {\bibfnamefont {K.}~\bibnamefont {Kargeti}}, \bibinfo {author}
  {\bibfnamefont {A.}~\bibnamefont {Jain}}, \bibinfo {author} {\bibfnamefont
  {R.}~\bibnamefont {Kumar}}, \bibinfo {author} {\bibfnamefont {S.~K.}\
  \bibnamefont {Panda}}, \bibinfo {author} {\bibfnamefont {S.~M.}\ \bibnamefont
  {Yusuf}}, \bibinfo {author} {\bibfnamefont {P.~L.}\ \bibnamefont {Paulose}},
  \bibinfo {author} {\bibfnamefont {U.~K.}\ \bibnamefont {Voma}}, \bibinfo
  {author} {\bibfnamefont {E.}~\bibnamefont {Kermarrec}}, \bibinfo {author}
  {\bibfnamefont {K.~H.}\ \bibnamefont {Kim}},\ and\ \bibinfo {author}
  {\bibfnamefont {B.}~\bibnamefont {Koteswararao}},\ }\bibfield  {title}
  {\bibinfo {title} {{Signatures of spin-liquid state in a 3D frustrated
  lattice compound \text{K}\text{Sr}\text{Fe$_{2}$}\text{(PO$_{4}$)$_{3}$} with
  S = 5/2}},\ }\href {https://doi.org/10.1063/5.0096942} {\bibfield  {journal}
  {\bibinfo  {journal} {APL Materials}\ }\textbf {\bibinfo {volume} {10}},\
  \bibinfo {pages} {101103} (\bibinfo {year} {2022})}\BibitemShut {NoStop}%
\bibitem [{\citenamefont {Li}\ \emph {et~al.}(2024)\citenamefont {Li},
  \citenamefont {Biswas},\ and\ \citenamefont
  {Parameswaran}}]{li2024classificationspin12fermionicquantum}%
  \BibitemOpen
  \bibfield  {author} {\bibinfo {author} {\bibfnamefont {M.-H.}\ \bibnamefont
  {Li}}, \bibinfo {author} {\bibfnamefont {S.}~\bibnamefont {Biswas}},\ and\
  \bibinfo {author} {\bibfnamefont {S.~A.}\ \bibnamefont {Parameswaran}},\
  }\href {https://arxiv.org/abs/2409.02898} {\bibinfo {title} {Classification
  of spin-$1/2$ fermionic quantum spin liquids on the trillium lattice}}
  (\bibinfo {year} {2024}),\ \Eprint {https://arxiv.org/abs/2409.02898}
  {arXiv:2409.02898 [cond-mat.str-el]} \BibitemShut {NoStop}%
\bibitem [{\citenamefont {Gonzalez}\ \emph {et~al.}(2024)\citenamefont
  {Gonzalez}, \citenamefont {Noculak}, \citenamefont {Sharma}, \citenamefont
  {Favre}, \citenamefont {Soh}, \citenamefont {Magrez}, \citenamefont {Bewley},
  \citenamefont {Jeschke}, \citenamefont {Reuther}, \citenamefont {R{\o}nnow},
  \citenamefont {Iqbal},\ and\ \citenamefont
  {{\v{Z}}ivkovi{\'{c}}}}]{Gonzalez2024}%
  \BibitemOpen
  \bibfield  {author} {\bibinfo {author} {\bibfnamefont {M.~G.}\ \bibnamefont
  {Gonzalez}}, \bibinfo {author} {\bibfnamefont {V.}~\bibnamefont {Noculak}},
  \bibinfo {author} {\bibfnamefont {A.}~\bibnamefont {Sharma}}, \bibinfo
  {author} {\bibfnamefont {V.}~\bibnamefont {Favre}}, \bibinfo {author}
  {\bibfnamefont {J.-R.}\ \bibnamefont {Soh}}, \bibinfo {author} {\bibfnamefont
  {A.}~\bibnamefont {Magrez}}, \bibinfo {author} {\bibfnamefont
  {R.}~\bibnamefont {Bewley}}, \bibinfo {author} {\bibfnamefont {H.~O.}\
  \bibnamefont {Jeschke}}, \bibinfo {author} {\bibfnamefont {J.}~\bibnamefont
  {Reuther}}, \bibinfo {author} {\bibfnamefont {H.~M.}\ \bibnamefont
  {R{\o}nnow}}, \bibinfo {author} {\bibfnamefont {Y.}~\bibnamefont {Iqbal}},\
  and\ \bibinfo {author} {\bibfnamefont {I.}~\bibnamefont
  {{\v{Z}}ivkovi{\'{c}}}},\ }\bibfield  {title} {\bibinfo {title} {Dynamics of
  \text{K$_{2}$}\text{Ni$_{2}$}(\text{SO$_{4})_{3}$} governed by proximity to a
  3\text{D} spin liquid model},\ }\href
  {https://doi.org/10.1038/s41467-024-51362-1} {\bibfield  {journal} {\bibinfo
  {journal} {Nat. Commun.}\ }\textbf {\bibinfo {volume} {15}},\ \bibinfo
  {pages} {7191} (\bibinfo {year} {2024})}\BibitemShut {NoStop}%
\bibitem [{\citenamefont {Khatua}\ \emph
  {et~al.}(2024{\natexlab{a}})\citenamefont {Khatua}, \citenamefont {Lee},
  \citenamefont {Ban}, \citenamefont {Uhlarz}, \citenamefont {Murugan},
  \citenamefont {Sankar}, \citenamefont {Hitti}, \citenamefont {Morris},
  \citenamefont {Choi},\ and\ \citenamefont {Khuntia}}]{PhysRevB.109.184432}%
  \BibitemOpen
  \bibfield  {author} {\bibinfo {author} {\bibfnamefont {J.}~\bibnamefont
  {Khatua}}, \bibinfo {author} {\bibfnamefont {S.}~\bibnamefont {Lee}},
  \bibinfo {author} {\bibfnamefont {G.}~\bibnamefont {Ban}}, \bibinfo {author}
  {\bibfnamefont {M.}~\bibnamefont {Uhlarz}}, \bibinfo {author} {\bibfnamefont
  {G.~S.}\ \bibnamefont {Murugan}}, \bibinfo {author} {\bibfnamefont
  {R.}~\bibnamefont {Sankar}}, \bibinfo {author} {\bibfnamefont
  {B.}~\bibnamefont {Hitti}}, \bibinfo {author} {\bibfnamefont
  {G.}~\bibnamefont {Morris}}, \bibinfo {author} {\bibfnamefont {K.-Y.}\
  \bibnamefont {Choi}},\ and\ \bibinfo {author} {\bibfnamefont
  {P.}~\bibnamefont {Khuntia}},\ }\bibfield  {title} {\bibinfo {title}
  {Magnetism and spin dynamics of the $s$=$\frac{3}{2}$ frustrated trillium
  lattice compound \text{K$_{2}$}\text{Cr}\text{Ti}\text{(PO$_{4}$)$_{3}$}},\
  }\href {https://doi.org/10.1103/PhysRevB.109.184432} {\bibfield  {journal}
  {\bibinfo  {journal} {Phys. Rev. B}\ }\textbf {\bibinfo {volume} {109}},\
  \bibinfo {pages} {184432} (\bibinfo {year} {2024}{\natexlab{a}})}\BibitemShut
  {NoStop}%
\bibitem [{\citenamefont {Yao}\ \emph {et~al.}(2023)\citenamefont {Yao},
  \citenamefont {Huang}, \citenamefont {Xie}, \citenamefont {Podlesnyak},
  \citenamefont {Brassington}, \citenamefont {Xing}, \citenamefont
  {Mudiyanselage}, \citenamefont {Wang}, \citenamefont {Xie}, \citenamefont
  {Zhang}, \citenamefont {Lee}, \citenamefont {Zapf}, \citenamefont {Bai},
  \citenamefont {Tennant}, \citenamefont {Liu},\ and\ \citenamefont
  {Zhou}}]{PhysRevLett.131.146701}%
  \BibitemOpen
  \bibfield  {author} {\bibinfo {author} {\bibfnamefont {W.}~\bibnamefont
  {Yao}}, \bibinfo {author} {\bibfnamefont {Q.}~\bibnamefont {Huang}}, \bibinfo
  {author} {\bibfnamefont {T.}~\bibnamefont {Xie}}, \bibinfo {author}
  {\bibfnamefont {A.}~\bibnamefont {Podlesnyak}}, \bibinfo {author}
  {\bibfnamefont {A.}~\bibnamefont {Brassington}}, \bibinfo {author}
  {\bibfnamefont {C.}~\bibnamefont {Xing}}, \bibinfo {author} {\bibfnamefont
  {R.~S.~D.}\ \bibnamefont {Mudiyanselage}}, \bibinfo {author} {\bibfnamefont
  {H.}~\bibnamefont {Wang}}, \bibinfo {author} {\bibfnamefont {W.}~\bibnamefont
  {Xie}}, \bibinfo {author} {\bibfnamefont {S.}~\bibnamefont {Zhang}}, \bibinfo
  {author} {\bibfnamefont {M.}~\bibnamefont {Lee}}, \bibinfo {author}
  {\bibfnamefont {V.~S.}\ \bibnamefont {Zapf}}, \bibinfo {author}
  {\bibfnamefont {X.}~\bibnamefont {Bai}}, \bibinfo {author} {\bibfnamefont
  {D.~A.}\ \bibnamefont {Tennant}}, \bibinfo {author} {\bibfnamefont
  {J.}~\bibnamefont {Liu}},\ and\ \bibinfo {author} {\bibfnamefont
  {H.}~\bibnamefont {Zhou}},\ }\bibfield  {title} {\bibinfo {title} {Continuous
  spin excitations in the three-dimensional frustrated magnet
  \text{K$_{2}$}\text{Ni$_{2}$}(\text{SO$_{4}$})$_{3}$},\ }\href
  {https://doi.org/10.1103/PhysRevLett.131.146701} {\bibfield  {journal}
  {\bibinfo  {journal} {Phys. Rev. Lett.}\ }\textbf {\bibinfo {volume} {131}},\
  \bibinfo {pages} {146701} (\bibinfo {year} {2023})}\BibitemShut {NoStop}%
\bibitem [{\citenamefont {Kub{\'i}{\v{c}}kov{\'a}}\ \emph
  {et~al.}(2024)\citenamefont {Kub{\'i}{\v{c}}kov{\'a}}, \citenamefont {Weber},
  \citenamefont {Panth\"{o}fer}, \citenamefont {Calder},\ and\ \citenamefont
  {M\"{o}ller}}]{Kub2024}%
  \BibitemOpen
  \bibfield  {author} {\bibinfo {author} {\bibfnamefont {L.}~\bibnamefont
  {Kub{\'i}{\v{c}}kov{\'a}}}, \bibinfo {author} {\bibfnamefont {A.~K.}\
  \bibnamefont {Weber}}, \bibinfo {author} {\bibfnamefont {M.}~\bibnamefont
  {Panth\"{o}fer}}, \bibinfo {author} {\bibfnamefont {S.}~\bibnamefont
  {Calder}},\ and\ \bibinfo {author} {\bibfnamefont {A.}~\bibnamefont
  {M\"{o}ller}},\ }\bibfield  {title} {\bibinfo {title}
  {\text{Cs$_{2}$}\text{Fe$_{2}$}(\text{MoO$_{4}$})$_{3}$ a strongly frustrated
  magnet with orbital degrees of freedom and magnetocaloric properties},\
  }\href {https://doi.org/10.1021/acs.chemmater.4c01262} {\bibfield  {journal}
  {\bibinfo  {journal} {Chem. Mat.}\ }\textbf {\bibinfo {volume} {36}},\
  \bibinfo {pages} {7016} (\bibinfo {year} {2024})}\BibitemShut {NoStop}%
\bibitem [{\citenamefont {Kolay}\ \emph {et~al.}(2024)\citenamefont {Kolay},
  \citenamefont {Ding}, \citenamefont {Furukawa}, \citenamefont {Tsirlin},\
  and\ \citenamefont {Nath}}]{PhysRevB.110.224405}%
  \BibitemOpen
  \bibfield  {author} {\bibinfo {author} {\bibfnamefont {R.}~\bibnamefont
  {Kolay}}, \bibinfo {author} {\bibfnamefont {Q.-P.}\ \bibnamefont {Ding}},
  \bibinfo {author} {\bibfnamefont {Y.}~\bibnamefont {Furukawa}}, \bibinfo
  {author} {\bibfnamefont {A.~A.}\ \bibnamefont {Tsirlin}},\ and\ \bibinfo
  {author} {\bibfnamefont {R.}~\bibnamefont {Nath}},\ }\bibfield  {title}
  {\bibinfo {title} {Magnetic properties of the double trillium lattice
  antiferromagnet \text{K}\text{Ba}\text{Cr$_{2}$}(\text{PO$_{4}$})$_{3}$},\
  }\href {https://doi.org/10.1103/PhysRevB.110.224405} {\bibfield  {journal}
  {\bibinfo  {journal} {Phys. Rev. B}\ }\textbf {\bibinfo {volume} {110}},\
  \bibinfo {pages} {224405} (\bibinfo {year} {2024})}\BibitemShut {NoStop}%
\bibitem [{\citenamefont {Mühlbauer}\ \emph {et~al.}(2009)\citenamefont
  {Mühlbauer}, \citenamefont {Binz}, \citenamefont {Jonietz}, \citenamefont
  {Pfleiderer}, \citenamefont {Rosch}, \citenamefont {Neubauer}, \citenamefont
  {Georgii},\ and\ \citenamefont {Böni}}]{science1166767}%
  \BibitemOpen
  \bibfield  {author} {\bibinfo {author} {\bibfnamefont {S.}~\bibnamefont
  {Mühlbauer}}, \bibinfo {author} {\bibfnamefont {B.}~\bibnamefont {Binz}},
  \bibinfo {author} {\bibfnamefont {F.}~\bibnamefont {Jonietz}}, \bibinfo
  {author} {\bibfnamefont {C.}~\bibnamefont {Pfleiderer}}, \bibinfo {author}
  {\bibfnamefont {A.}~\bibnamefont {Rosch}}, \bibinfo {author} {\bibfnamefont
  {A.}~\bibnamefont {Neubauer}}, \bibinfo {author} {\bibfnamefont
  {R.}~\bibnamefont {Georgii}},\ and\ \bibinfo {author} {\bibfnamefont
  {P.}~\bibnamefont {Böni}},\ }\bibfield  {title} {\bibinfo {title} {Skyrmion
  lattice in a chiral magnet},\ }\href
  {https://doi.org/10.1126/science.1166767} {\bibfield  {journal} {\bibinfo
  {journal} {Science}\ }\textbf {\bibinfo {volume} {323}},\ \bibinfo {pages}
  {915} (\bibinfo {year} {2009})}\BibitemShut {NoStop}%
\bibitem [{\citenamefont {Kakihana}\ \emph {et~al.}(2017)\citenamefont
  {Kakihana}, \citenamefont {Nishimura}, \citenamefont {Ashitomi},
  \citenamefont {Yara}, \citenamefont {Aoki}, \citenamefont {Nakamura},
  \citenamefont {Honda}, \citenamefont {Nakashima}, \citenamefont {Amako},
  \citenamefont {Uwatoko}, \citenamefont {Sakakibara}, \citenamefont
  {Nakamura}, \citenamefont {Takeuchi}, \citenamefont {Haga}, \citenamefont
  {Yamamoto}, \citenamefont {Harima}, \citenamefont {Hedo}, \citenamefont
  {Nakama},\ and\ \citenamefont {{\={O}}nuki}}]{Kakihana2017}%
  \BibitemOpen
  \bibfield  {author} {\bibinfo {author} {\bibfnamefont {M.}~\bibnamefont
  {Kakihana}}, \bibinfo {author} {\bibfnamefont {K.}~\bibnamefont {Nishimura}},
  \bibinfo {author} {\bibfnamefont {Y.}~\bibnamefont {Ashitomi}}, \bibinfo
  {author} {\bibfnamefont {T.}~\bibnamefont {Yara}}, \bibinfo {author}
  {\bibfnamefont {D.}~\bibnamefont {Aoki}}, \bibinfo {author} {\bibfnamefont
  {A.}~\bibnamefont {Nakamura}}, \bibinfo {author} {\bibfnamefont
  {F.}~\bibnamefont {Honda}}, \bibinfo {author} {\bibfnamefont
  {M.}~\bibnamefont {Nakashima}}, \bibinfo {author} {\bibfnamefont
  {Y.}~\bibnamefont {Amako}}, \bibinfo {author} {\bibfnamefont
  {Y.}~\bibnamefont {Uwatoko}}, \bibinfo {author} {\bibfnamefont
  {T.}~\bibnamefont {Sakakibara}}, \bibinfo {author} {\bibfnamefont
  {S.}~\bibnamefont {Nakamura}}, \bibinfo {author} {\bibfnamefont
  {T.}~\bibnamefont {Takeuchi}}, \bibinfo {author} {\bibfnamefont
  {Y.}~\bibnamefont {Haga}}, \bibinfo {author} {\bibfnamefont {E.}~\bibnamefont
  {Yamamoto}}, \bibinfo {author} {\bibfnamefont {H.}~\bibnamefont {Harima}},
  \bibinfo {author} {\bibfnamefont {M.}~\bibnamefont {Hedo}}, \bibinfo {author}
  {\bibfnamefont {T.}~\bibnamefont {Nakama}},\ and\ \bibinfo {author}
  {\bibfnamefont {Y.}~\bibnamefont {{\={O}}nuki}},\ }\bibfield  {title}
  {\bibinfo {title} {Unique electronic states in non-centrosymmetric cubic
  compounds},\ }\href {https://doi.org/10.1007/s11664-016-5265-z} {\bibfield
  {journal} {\bibinfo  {journal} {J. Electron. Mater.}\ }\textbf {\bibinfo
  {volume} {46}},\ \bibinfo {pages} {3572} (\bibinfo {year}
  {2017})}\BibitemShut {NoStop}%
\bibitem [{\citenamefont {Mahraj}\ and\ \citenamefont
  {Ptok}(2025)}]{mahraj2024chiralphononicelectronicedge}%
  \BibitemOpen
  \bibfield  {author} {\bibinfo {author} {\bibfnamefont {I.}~\bibnamefont
  {Mahraj}}\ and\ \bibinfo {author} {\bibfnamefont {A.}~\bibnamefont {Ptok}},\
  }\bibfield  {title} {\bibinfo {title} {Chiral phononic and electronic edge
  modes of \text{Eu}\text{Pt}\text{Si}},\ }\href
  {https://doi.org/10.1103/PhysRevB.111.165132} {\bibfield  {journal} {\bibinfo
   {journal} {Phys. Rev. B}\ }\textbf {\bibinfo {volume} {111}},\ \bibinfo
  {pages} {165132} (\bibinfo {year} {2025})}\BibitemShut {NoStop}%
\bibitem [{\citenamefont {Redpath}\ and\ \citenamefont
  {Hopkinson}(2010)}]{PhysRevB.82.014410}%
  \BibitemOpen
  \bibfield  {author} {\bibinfo {author} {\bibfnamefont {T.~E.}\ \bibnamefont
  {Redpath}}\ and\ \bibinfo {author} {\bibfnamefont {J.~M.}\ \bibnamefont
  {Hopkinson}},\ }\bibfield  {title} {\bibinfo {title} {Spin ice on the
  trillium lattice studied by monte carlo calculations},\ }\href
  {https://doi.org/10.1103/PhysRevB.82.014410} {\bibfield  {journal} {\bibinfo
  {journal} {Phys. Rev. B}\ }\textbf {\bibinfo {volume} {82}},\ \bibinfo
  {pages} {014410} (\bibinfo {year} {2010})}\BibitemShut {NoStop}%
\bibitem [{\citenamefont {Isakov}\ \emph {et~al.}(2008)\citenamefont {Isakov},
  \citenamefont {Hopkinson},\ and\ \citenamefont {Kee}}]{PhysRevB.78.014404}%
  \BibitemOpen
  \bibfield  {author} {\bibinfo {author} {\bibfnamefont {S.~V.}\ \bibnamefont
  {Isakov}}, \bibinfo {author} {\bibfnamefont {J.~M.}\ \bibnamefont
  {Hopkinson}},\ and\ \bibinfo {author} {\bibfnamefont {H.-Y.}\ \bibnamefont
  {Kee}},\ }\bibfield  {title} {\bibinfo {title} {Fate of partial order on
  trillium and distorted windmill lattices},\ }\href
  {https://doi.org/10.1103/PhysRevB.78.014404} {\bibfield  {journal} {\bibinfo
  {journal} {Phys. Rev. B}\ }\textbf {\bibinfo {volume} {78}},\ \bibinfo
  {pages} {014404} (\bibinfo {year} {2008})}\BibitemShut {NoStop}%
\bibitem [{\citenamefont {Bulled}\ \emph {et~al.}(2022)\citenamefont {Bulled},
  \citenamefont {Paddison}, \citenamefont {Wildes}, \citenamefont {Lhotel},
  \citenamefont {Cassidy}, \citenamefont {Pato-Dold\'an}, \citenamefont
  {G\'omez-Aguirre}, \citenamefont {Saines},\ and\ \citenamefont
  {Goodwin}}]{PhysRevLett.128.177201}%
  \BibitemOpen
  \bibfield  {author} {\bibinfo {author} {\bibfnamefont {J.~M.}\ \bibnamefont
  {Bulled}}, \bibinfo {author} {\bibfnamefont {J.~A.~M.}\ \bibnamefont
  {Paddison}}, \bibinfo {author} {\bibfnamefont {A.}~\bibnamefont {Wildes}},
  \bibinfo {author} {\bibfnamefont {E.}~\bibnamefont {Lhotel}}, \bibinfo
  {author} {\bibfnamefont {S.~J.}\ \bibnamefont {Cassidy}}, \bibinfo {author}
  {\bibfnamefont {B.}~\bibnamefont {Pato-Dold\'an}}, \bibinfo {author}
  {\bibfnamefont {L.~C.}\ \bibnamefont {G\'omez-Aguirre}}, \bibinfo {author}
  {\bibfnamefont {P.~J.}\ \bibnamefont {Saines}},\ and\ \bibinfo {author}
  {\bibfnamefont {A.~L.}\ \bibnamefont {Goodwin}},\ }\bibfield  {title}
  {\bibinfo {title} {Geometric frustration on the trillium lattice in a
  magnetic metal-organic framework},\ }\href
  {https://doi.org/10.1103/PhysRevLett.128.177201} {\bibfield  {journal}
  {\bibinfo  {journal} {Phys. Rev. Lett.}\ }\textbf {\bibinfo {volume} {128}},\
  \bibinfo {pages} {177201} (\bibinfo {year} {2022})}\BibitemShut {NoStop}%
\bibitem [{\citenamefont {Zatovsky}\ \emph {et~al.}(2007)\citenamefont
  {Zatovsky}, \citenamefont {Yatskin}, \citenamefont {Baumer}, \citenamefont
  {Slobodyanik},\ and\ \citenamefont {Shishkin}}]{Zatovskywm2157}%
  \BibitemOpen
  \bibfield  {author} {\bibinfo {author} {\bibfnamefont {I.~V.}\ \bibnamefont
  {Zatovsky}}, \bibinfo {author} {\bibfnamefont {M.~M.}\ \bibnamefont
  {Yatskin}}, \bibinfo {author} {\bibfnamefont {V.~N.}\ \bibnamefont {Baumer}},
  \bibinfo {author} {\bibfnamefont {N.~S.}\ \bibnamefont {Slobodyanik}},\ and\
  \bibinfo {author} {\bibfnamefont {O.~V.}\ \bibnamefont {Shishkin}},\
  }\bibfield  {title} {\bibinfo {title} {{Langbeinite-related
  \text{K$_{2}$}\text{FeSn}\text{(PO$_{4}$)$_{3}$} from single-crystal data}},\
  }\href {https://doi.org/10.1107/S1600536807058588} {\bibfield  {journal}
  {\bibinfo  {journal} {Acta crystallogr. Section E}\ }\textbf {\bibinfo
  {volume} {63}},\ \bibinfo {pages} {i199} (\bibinfo {year}
  {2007})}\BibitemShut {NoStop}%
\bibitem [{sm()}]{sm}%
  \BibitemOpen
  \href@noop {} {}\bibinfo {note} {See supplementary material for further
  details on sample synthesis, crystal structure, specific heat, and muon spin
  relaxation data analysis}\BibitemShut {NoStop}%
\bibitem [{\citenamefont {Suter}\ and\ \citenamefont
  {Wojek}(2012)}]{SUTER201269}%
  \BibitemOpen
  \bibfield  {author} {\bibinfo {author} {\bibfnamefont {A.}~\bibnamefont
  {Suter}}\ and\ \bibinfo {author} {\bibfnamefont {B.}~\bibnamefont {Wojek}},\
  }\bibfield  {title} {\bibinfo {title} {Musrfit: A free platform-independent
  framework for $\mu$\text{S}\text{R} data analysis},\ }\href
  {https://doi.org/https://doi.org/10.1016/j.phpro.2012.04.042} {\bibfield
  {journal} {\bibinfo  {journal} {Phys. Procedia}\ }\textbf {\bibinfo {volume}
  {30}},\ \bibinfo {pages} {69} (\bibinfo {year} {2012})}\BibitemShut {NoStop}%
\bibitem [{\citenamefont {Toby}(2001)}]{Tobyhw0089}%
  \BibitemOpen
  \bibfield  {author} {\bibinfo {author} {\bibfnamefont {B.~H.}\ \bibnamefont
  {Toby}},\ }\bibfield  {title} {\bibinfo {title} {{{\it EXPGUI}, a graphical
  user interface for {\it GSAS}}},\ }\href
  {https://doi.org/10.1107/S0021889801002242} {\bibfield  {journal} {\bibinfo
  {journal} {J. Appl. Crystallogr.}\ }\textbf {\bibinfo {volume} {34}},\
  \bibinfo {pages} {210} (\bibinfo {year} {2001})}\BibitemShut {NoStop}%
\bibitem [{\citenamefont {Sen}\ \emph {et~al.}(2012)\citenamefont {Sen},
  \citenamefont {Damle},\ and\ \citenamefont {Moessner}}]{PhysRevB.86.205134}%
  \BibitemOpen
  \bibfield  {author} {\bibinfo {author} {\bibfnamefont {A.}~\bibnamefont
  {Sen}}, \bibinfo {author} {\bibfnamefont {K.}~\bibnamefont {Damle}},\ and\
  \bibinfo {author} {\bibfnamefont {R.}~\bibnamefont {Moessner}},\ }\bibfield
  {title} {\bibinfo {title} {Vacancy-induced spin textures and their
  interactions in a classical spin liquid},\ }\href
  {https://doi.org/10.1103/PhysRevB.86.205134} {\bibfield  {journal} {\bibinfo
  {journal} {Phys. Rev. B}\ }\textbf {\bibinfo {volume} {86}},\ \bibinfo
  {pages} {205134} (\bibinfo {year} {2012})}\BibitemShut {NoStop}%
\bibitem [{\citenamefont {Sibille}\ \emph {et~al.}(2017)\citenamefont
  {Sibille}, \citenamefont {Lhotel}, \citenamefont {Ciomaga~Hatnean},
  \citenamefont {Nilsen}, \citenamefont {Ehlers}, \citenamefont {Cervellino},
  \citenamefont {Ressouche}, \citenamefont {Frontzek}, \citenamefont {Zaharko},
  \citenamefont {Pomjakushin}, \citenamefont {Stuhr}, \citenamefont {Walker},
  \citenamefont {Adroja}, \citenamefont {Luetkens}, \citenamefont {Baines},
  \citenamefont {Amato}, \citenamefont {Balakrishnan}, \citenamefont
  {Fennell},\ and\ \citenamefont {Kenzelmann}}]{Sibille2017}%
  \BibitemOpen
  \bibfield  {author} {\bibinfo {author} {\bibfnamefont {R.}~\bibnamefont
  {Sibille}}, \bibinfo {author} {\bibfnamefont {E.}~\bibnamefont {Lhotel}},
  \bibinfo {author} {\bibfnamefont {M.}~\bibnamefont {Ciomaga~Hatnean}},
  \bibinfo {author} {\bibfnamefont {G.~J.}\ \bibnamefont {Nilsen}}, \bibinfo
  {author} {\bibfnamefont {G.}~\bibnamefont {Ehlers}}, \bibinfo {author}
  {\bibfnamefont {A.}~\bibnamefont {Cervellino}}, \bibinfo {author}
  {\bibfnamefont {E.}~\bibnamefont {Ressouche}}, \bibinfo {author}
  {\bibfnamefont {M.}~\bibnamefont {Frontzek}}, \bibinfo {author}
  {\bibfnamefont {O.}~\bibnamefont {Zaharko}}, \bibinfo {author} {\bibfnamefont
  {V.}~\bibnamefont {Pomjakushin}}, \bibinfo {author} {\bibfnamefont
  {U.}~\bibnamefont {Stuhr}}, \bibinfo {author} {\bibfnamefont {H.~C.}\
  \bibnamefont {Walker}}, \bibinfo {author} {\bibfnamefont {D.~T.}\
  \bibnamefont {Adroja}}, \bibinfo {author} {\bibfnamefont {H.}~\bibnamefont
  {Luetkens}}, \bibinfo {author} {\bibfnamefont {C.}~\bibnamefont {Baines}},
  \bibinfo {author} {\bibfnamefont {A.}~\bibnamefont {Amato}}, \bibinfo
  {author} {\bibfnamefont {G.}~\bibnamefont {Balakrishnan}}, \bibinfo {author}
  {\bibfnamefont {T.}~\bibnamefont {Fennell}},\ and\ \bibinfo {author}
  {\bibfnamefont {M.}~\bibnamefont {Kenzelmann}},\ }\bibfield  {title}
  {\bibinfo {title} {Coulomb spin liquid in anion-disordered pyrochlore
  \text{Tb$_{2}$}\text{Hf$_{2}$}\text{O$_{7}$}},\ }\href
  {https://doi.org/10.1038/s41467-017-00905-w} {\bibfield  {journal} {\bibinfo
  {journal} {Nat. Commun.}\ }\textbf {\bibinfo {volume} {8}},\ \bibinfo {pages}
  {892} (\bibinfo {year} {2017})}\BibitemShut {NoStop}%
\bibitem [{\citenamefont {Moriya}(1960)}]{PhysRev.120.91}%
  \BibitemOpen
  \bibfield  {author} {\bibinfo {author} {\bibfnamefont {T.}~\bibnamefont
  {Moriya}},\ }\bibfield  {title} {\bibinfo {title} {Anisotropic superexchange
  interaction and weak ferromagnetism},\ }\href
  {https://doi.org/10.1103/PhysRev.120.91} {\bibfield  {journal} {\bibinfo
  {journal} {Phys. Rev.}\ }\textbf {\bibinfo {volume} {120}},\ \bibinfo {pages}
  {91} (\bibinfo {year} {1960})}\BibitemShut {NoStop}%
\bibitem [{\citenamefont {Svoboda}\ \emph {et~al.}(2001)\citenamefont
  {Svoboda}, \citenamefont {Javorsk\'y}, \citenamefont
  {Divi\ifmmode~\check{s}\else \v{s}\fi{}}, \citenamefont {Sechovsk\'y},
  \citenamefont {Honda}, \citenamefont {Oomi},\ and\ \citenamefont
  {Menovsky}}]{PhysRevB.63.212408}%
  \BibitemOpen
  \bibfield  {author} {\bibinfo {author} {\bibfnamefont {P.}~\bibnamefont
  {Svoboda}}, \bibinfo {author} {\bibfnamefont {P.}~\bibnamefont {Javorsk\'y}},
  \bibinfo {author} {\bibfnamefont {M.}~\bibnamefont
  {Divi\ifmmode~\check{s}\else \v{s}\fi{}}}, \bibinfo {author} {\bibfnamefont
  {V.}~\bibnamefont {Sechovsk\'y}}, \bibinfo {author} {\bibfnamefont
  {F.}~\bibnamefont {Honda}}, \bibinfo {author} {\bibfnamefont
  {G.}~\bibnamefont {Oomi}},\ and\ \bibinfo {author} {\bibfnamefont {A.~A.}\
  \bibnamefont {Menovsky}},\ }\bibfield  {title} {\bibinfo {title} {Importance
  of anharmonic terms in the analysis of the specific heat of
  \text{U}\text{Ni$_{2}$}\text{Si$_{2}$}},\ }\href
  {https://doi.org/10.1103/PhysRevB.63.212408} {\bibfield  {journal} {\bibinfo
  {journal} {Phys. Rev. B}\ }\textbf {\bibinfo {volume} {63}},\ \bibinfo
  {pages} {212408} (\bibinfo {year} {2001})}\BibitemShut {NoStop}%
\bibitem [{\citenamefont {Khatua}\ \emph
  {et~al.}(2024{\natexlab{b}})\citenamefont {Khatua}, \citenamefont
  {Gomil\ifmmode~\check{s}\else \v{s}\fi{}ek}, \citenamefont {Choi},\ and\
  \citenamefont {Khuntia}}]{PhysRevB.110.184402}%
  \BibitemOpen
  \bibfield  {author} {\bibinfo {author} {\bibfnamefont {J.}~\bibnamefont
  {Khatua}}, \bibinfo {author} {\bibfnamefont {M.}~\bibnamefont
  {Gomil\ifmmode~\check{s}\else \v{s}\fi{}ek}}, \bibinfo {author}
  {\bibfnamefont {K.-Y.}\ \bibnamefont {Choi}},\ and\ \bibinfo {author}
  {\bibfnamefont {P.}~\bibnamefont {Khuntia}},\ }\bibfield  {title} {\bibinfo
  {title} {Magnetism and field-induced effects in the s = 5/2 honeycomb lattice
  antiferromagnet \text{Fe}\text{P$_{3}$}\text{Si}\text{O$_{11}$}},\ }\href
  {https://doi.org/10.1103/PhysRevB.110.184402} {\bibfield  {journal} {\bibinfo
   {journal} {Phys. Rev. B}\ }\textbf {\bibinfo {volume} {110}},\ \bibinfo
  {pages} {184402} (\bibinfo {year} {2024}{\natexlab{b}})}\BibitemShut
  {NoStop}%
\bibitem [{\citenamefont {Ramirez}\ \emph {et~al.}(2000)\citenamefont
  {Ramirez}, \citenamefont {Hessen},\ and\ \citenamefont
  {Winklemann}}]{PhysRevLett.84.2957}%
  \BibitemOpen
  \bibfield  {author} {\bibinfo {author} {\bibfnamefont {A.~P.}\ \bibnamefont
  {Ramirez}}, \bibinfo {author} {\bibfnamefont {B.}~\bibnamefont {Hessen}},\
  and\ \bibinfo {author} {\bibfnamefont {M.}~\bibnamefont {Winklemann}},\
  }\bibfield  {title} {\bibinfo {title} {Entropy balance and evidence for local
  spin singlets in a kagom\'e-like magnet},\ }\href
  {https://doi.org/10.1103/PhysRevLett.84.2957} {\bibfield  {journal} {\bibinfo
   {journal} {Phys. Rev. Lett.}\ }\textbf {\bibinfo {volume} {84}},\ \bibinfo
  {pages} {2957} (\bibinfo {year} {2000})}\BibitemShut {NoStop}%
\bibitem [{\citenamefont {Khatua}\ \emph {et~al.}(2022)\citenamefont {Khatua},
  \citenamefont {Gomil{\v{s}}ek}, \citenamefont {Orain}, \citenamefont
  {Strydom}, \citenamefont {Jagli{\v{c}}i{\'{c}}}, \citenamefont {Colin},
  \citenamefont {Petit}, \citenamefont {Ozarowski}, \citenamefont
  {Mangin-Thro}, \citenamefont {Sethupathi}, \citenamefont {Rao}, \citenamefont
  {Zorko},\ and\ \citenamefont {Khuntia}}]{Khatua2022}%
  \BibitemOpen
  \bibfield  {author} {\bibinfo {author} {\bibfnamefont {J.}~\bibnamefont
  {Khatua}}, \bibinfo {author} {\bibfnamefont {M.}~\bibnamefont
  {Gomil{\v{s}}ek}}, \bibinfo {author} {\bibfnamefont {J.~C.}\ \bibnamefont
  {Orain}}, \bibinfo {author} {\bibfnamefont {A.~M.}\ \bibnamefont {Strydom}},
  \bibinfo {author} {\bibfnamefont {Z.}~\bibnamefont {Jagli{\v{c}}i{\'{c}}}},
  \bibinfo {author} {\bibfnamefont {C.~V.}\ \bibnamefont {Colin}}, \bibinfo
  {author} {\bibfnamefont {S.}~\bibnamefont {Petit}}, \bibinfo {author}
  {\bibfnamefont {A.}~\bibnamefont {Ozarowski}}, \bibinfo {author}
  {\bibfnamefont {L.}~\bibnamefont {Mangin-Thro}}, \bibinfo {author}
  {\bibfnamefont {K.}~\bibnamefont {Sethupathi}}, \bibinfo {author}
  {\bibfnamefont {M.~S.~R.}\ \bibnamefont {Rao}}, \bibinfo {author}
  {\bibfnamefont {A.}~\bibnamefont {Zorko}},\ and\ \bibinfo {author}
  {\bibfnamefont {P.}~\bibnamefont {Khuntia}},\ }\bibfield  {title} {\bibinfo
  {title} {Signature of a randomness-driven spin-liquid state in a frustrated
  magnet},\ }\href {https://doi.org/10.1038/s42005-022-00879-2} {\bibfield
  {journal} {\bibinfo  {journal} {Communications Physics}\ }\textbf {\bibinfo
  {volume} {5}},\ \bibinfo {pages} {99} (\bibinfo {year} {2022})}\BibitemShut
  {NoStop}%
\bibitem [{\citenamefont {Lee}\ \emph {et~al.}(2016)\citenamefont {Lee},
  \citenamefont {Do}, \citenamefont {Lee}, \citenamefont {Choi}, \citenamefont
  {Lee}, \citenamefont {Choi}, \citenamefont {Reyes}, \citenamefont {Kuhns},
  \citenamefont {Ozarowski},\ and\ \citenamefont {Choi}}]{PhysRevB.93.174402}%
  \BibitemOpen
  \bibfield  {author} {\bibinfo {author} {\bibfnamefont {S.}~\bibnamefont
  {Lee}}, \bibinfo {author} {\bibfnamefont {S.-H.}\ \bibnamefont {Do}},
  \bibinfo {author} {\bibfnamefont {W.-J.}\ \bibnamefont {Lee}}, \bibinfo
  {author} {\bibfnamefont {Y.~S.}\ \bibnamefont {Choi}}, \bibinfo {author}
  {\bibfnamefont {M.}~\bibnamefont {Lee}}, \bibinfo {author} {\bibfnamefont
  {E.~S.}\ \bibnamefont {Choi}}, \bibinfo {author} {\bibfnamefont {A.~P.}\
  \bibnamefont {Reyes}}, \bibinfo {author} {\bibfnamefont {P.~L.}\ \bibnamefont
  {Kuhns}}, \bibinfo {author} {\bibfnamefont {A.}~\bibnamefont {Ozarowski}},\
  and\ \bibinfo {author} {\bibfnamefont {K.-Y.}\ \bibnamefont {Choi}},\
  }\bibfield  {title} {\bibinfo {title} {Multistage symmetry breaking in the
  breathing pyrochlore lattice
  \text{Li}\text{(Ga,In)}\text{Cr$_{4}$}\text{O$_{8}$}},\ }\href
  {https://doi.org/10.1103/PhysRevB.93.174402} {\bibfield  {journal} {\bibinfo
  {journal} {Phys. Rev. B}\ }\textbf {\bibinfo {volume} {93}},\ \bibinfo
  {pages} {174402} (\bibinfo {year} {2016})}\BibitemShut {NoStop}%
\bibitem [{\citenamefont {Lee}\ \emph {et~al.}(2022)\citenamefont {Lee},
  \citenamefont {Zhu}, \citenamefont {Oshima}, \citenamefont {Shiroka},
  \citenamefont {Wang}, \citenamefont {Luetkens}, \citenamefont {Yang},
  \citenamefont {L\"u},\ and\ \citenamefont {Choi}}]{PhysRevB.105.094439}%
  \BibitemOpen
  \bibfield  {author} {\bibinfo {author} {\bibfnamefont {S.}~\bibnamefont
  {Lee}}, \bibinfo {author} {\bibfnamefont {T.}~\bibnamefont {Zhu}}, \bibinfo
  {author} {\bibfnamefont {Y.}~\bibnamefont {Oshima}}, \bibinfo {author}
  {\bibfnamefont {T.}~\bibnamefont {Shiroka}}, \bibinfo {author} {\bibfnamefont
  {C.}~\bibnamefont {Wang}}, \bibinfo {author} {\bibfnamefont {H.}~\bibnamefont
  {Luetkens}}, \bibinfo {author} {\bibfnamefont {H.}~\bibnamefont {Yang}},
  \bibinfo {author} {\bibfnamefont {M.}~\bibnamefont {L\"u}},\ and\ \bibinfo
  {author} {\bibfnamefont {K.-Y.}\ \bibnamefont {Choi}},\ }\bibfield  {title}
  {\bibinfo {title} {Timescale distributions of spin fluctuations in the $s=2$
  kagome antiferromagnet
  \text{Cs}\text{Mn$_{3}$}\text{F$_{6}$}\text{(SeO$_{3}$)$_{2}$}},\ }\href
  {https://doi.org/10.1103/PhysRevB.105.094439} {\bibfield  {journal} {\bibinfo
   {journal} {Phys. Rev. B}\ }\textbf {\bibinfo {volume} {105}},\ \bibinfo
  {pages} {094439} (\bibinfo {year} {2022})}\BibitemShut {NoStop}%
\bibitem [{\citenamefont {Glamazda}\ \emph {et~al.}(2017)\citenamefont
  {Glamazda}, \citenamefont {Choi}, \citenamefont {Do}, \citenamefont {Lee},
  \citenamefont {Lemmens}, \citenamefont {Ponomaryov}, \citenamefont {Zvyagin},
  \citenamefont {Wosnitza}, \citenamefont {Sari}, \citenamefont {Watanabe},\
  and\ \citenamefont {Choi}}]{PhysRevB.95.184430}%
  \BibitemOpen
  \bibfield  {author} {\bibinfo {author} {\bibfnamefont {A.}~\bibnamefont
  {Glamazda}}, \bibinfo {author} {\bibfnamefont {Y.~S.}\ \bibnamefont {Choi}},
  \bibinfo {author} {\bibfnamefont {S.-H.}\ \bibnamefont {Do}}, \bibinfo
  {author} {\bibfnamefont {S.}~\bibnamefont {Lee}}, \bibinfo {author}
  {\bibfnamefont {P.}~\bibnamefont {Lemmens}}, \bibinfo {author} {\bibfnamefont
  {A.~N.}\ \bibnamefont {Ponomaryov}}, \bibinfo {author} {\bibfnamefont
  {S.~A.}\ \bibnamefont {Zvyagin}}, \bibinfo {author} {\bibfnamefont
  {J.}~\bibnamefont {Wosnitza}}, \bibinfo {author} {\bibfnamefont {D.~P.}\
  \bibnamefont {Sari}}, \bibinfo {author} {\bibfnamefont {I.}~\bibnamefont
  {Watanabe}},\ and\ \bibinfo {author} {\bibfnamefont {K.-Y.}\ \bibnamefont
  {Choi}},\ }\bibfield  {title} {\bibinfo {title} {Quantum criticality in the
  coupled two-leg spin ladder
  \text{Ba$_{2}$}\text{Cu}\text{Te}\text{O$_{6}$}},\ }\href
  {https://doi.org/10.1103/PhysRevB.95.184430} {\bibfield  {journal} {\bibinfo
  {journal} {Phys. Rev. B}\ }\textbf {\bibinfo {volume} {95}},\ \bibinfo
  {pages} {184430} (\bibinfo {year} {2017})}\BibitemShut {NoStop}%
\bibitem [{\citenamefont {Le~Yaouanc}\ and\ \citenamefont
  {De~Reotier}(2011)}]{le2011muon}%
  \BibitemOpen
  \bibfield  {author} {\bibinfo {author} {\bibfnamefont {A.}~\bibnamefont
  {Le~Yaouanc}}\ and\ \bibinfo {author} {\bibfnamefont {P.~D.}\ \bibnamefont
  {De~Reotier}},\ }\href@noop {} {\emph {\bibinfo {title} {Muon spin rotation,
  relaxation, and resonance: applications to condensed matter}}},\ \bibinfo
  {number} {147}\ (\bibinfo  {publisher} {OUP Oxford},\ \bibinfo {year}
  {2011})\BibitemShut {NoStop}%
\bibitem [{\citenamefont {Uemura}\ \emph {et~al.}(1985)\citenamefont {Uemura},
  \citenamefont {Yamazaki}, \citenamefont {Harshman}, \citenamefont {Senba},\
  and\ \citenamefont {Ansaldo}}]{PhysRevB.31.546}%
  \BibitemOpen
  \bibfield  {author} {\bibinfo {author} {\bibfnamefont {Y.~J.}\ \bibnamefont
  {Uemura}}, \bibinfo {author} {\bibfnamefont {T.}~\bibnamefont {Yamazaki}},
  \bibinfo {author} {\bibfnamefont {D.~R.}\ \bibnamefont {Harshman}}, \bibinfo
  {author} {\bibfnamefont {M.}~\bibnamefont {Senba}},\ and\ \bibinfo {author}
  {\bibfnamefont {E.~J.}\ \bibnamefont {Ansaldo}},\ }\bibfield  {title}
  {\bibinfo {title} {Muon-spin relaxation in \text{Au}\text{Fe} and
  \text{Cu}\text{Mn} spin glasses},\ }\href
  {https://doi.org/10.1103/PhysRevB.31.546} {\bibfield  {journal} {\bibinfo
  {journal} {Phys. Rev. B}\ }\textbf {\bibinfo {volume} {31}},\ \bibinfo
  {pages} {546} (\bibinfo {year} {1985})}\BibitemShut {NoStop}%
\bibitem [{\citenamefont {Cai}\ \emph {et~al.}(2018)\citenamefont {Cai},
  \citenamefont {Wilson}, \citenamefont {Hallas}, \citenamefont {Liu},
  \citenamefont {Frandsen}, \citenamefont {Dunsiger}, \citenamefont {Krizan},
  \citenamefont {Cava}, \citenamefont {Rubel}, \citenamefont {Uemura},\ and\
  \citenamefont {Luke}}]{Cai2018}%
  \BibitemOpen
  \bibfield  {author} {\bibinfo {author} {\bibfnamefont {Y.}~\bibnamefont
  {Cai}}, \bibinfo {author} {\bibfnamefont {M.~N.}\ \bibnamefont {Wilson}},
  \bibinfo {author} {\bibfnamefont {A.~M.}\ \bibnamefont {Hallas}}, \bibinfo
  {author} {\bibfnamefont {L.}~\bibnamefont {Liu}}, \bibinfo {author}
  {\bibfnamefont {B.~A.}\ \bibnamefont {Frandsen}}, \bibinfo {author}
  {\bibfnamefont {S.~R.}\ \bibnamefont {Dunsiger}}, \bibinfo {author}
  {\bibfnamefont {J.~W.}\ \bibnamefont {Krizan}}, \bibinfo {author}
  {\bibfnamefont {R.~J.}\ \bibnamefont {Cava}}, \bibinfo {author}
  {\bibfnamefont {O.}~\bibnamefont {Rubel}}, \bibinfo {author} {\bibfnamefont
  {Y.~J.}\ \bibnamefont {Uemura}},\ and\ \bibinfo {author} {\bibfnamefont
  {G.~M.}\ \bibnamefont {Luke}},\ }\bibfield  {title} {\bibinfo {title}
  {$\mu$\text{SR} study of spin freezing and persistent spin dynamics in
  \text{Na}\text{Ca}\text{Ni$_{2}$}\text{F$_{7}$}},\ }\href
  {https://doi.org/10.1088/1361-648X/aad91c} {\bibfield  {journal} {\bibinfo
  {journal} {J. Phys. Condens. Matter.}\ }\textbf {\bibinfo {volume} {30}},\
  \bibinfo {pages} {385802} (\bibinfo {year} {2018})}\BibitemShut {NoStop}%
\bibitem [{\citenamefont {Uemura}\ \emph {et~al.}(1994)\citenamefont {Uemura},
  \citenamefont {Keren}, \citenamefont {Kojima}, \citenamefont {Le},
  \citenamefont {Luke}, \citenamefont {Wu}, \citenamefont {Ajiro},
  \citenamefont {Asano}, \citenamefont {Kuriyama}, \citenamefont {Mekata},
  \citenamefont {Kikuchi},\ and\ \citenamefont
  {Kakurai}}]{PhysRevLett.73.3306}%
  \BibitemOpen
  \bibfield  {author} {\bibinfo {author} {\bibfnamefont {Y.~J.}\ \bibnamefont
  {Uemura}}, \bibinfo {author} {\bibfnamefont {A.}~\bibnamefont {Keren}},
  \bibinfo {author} {\bibfnamefont {K.}~\bibnamefont {Kojima}}, \bibinfo
  {author} {\bibfnamefont {L.~P.}\ \bibnamefont {Le}}, \bibinfo {author}
  {\bibfnamefont {G.~M.}\ \bibnamefont {Luke}}, \bibinfo {author}
  {\bibfnamefont {W.~D.}\ \bibnamefont {Wu}}, \bibinfo {author} {\bibfnamefont
  {Y.}~\bibnamefont {Ajiro}}, \bibinfo {author} {\bibfnamefont
  {T.}~\bibnamefont {Asano}}, \bibinfo {author} {\bibfnamefont
  {Y.}~\bibnamefont {Kuriyama}}, \bibinfo {author} {\bibfnamefont
  {M.}~\bibnamefont {Mekata}}, \bibinfo {author} {\bibfnamefont
  {H.}~\bibnamefont {Kikuchi}},\ and\ \bibinfo {author} {\bibfnamefont
  {K.}~\bibnamefont {Kakurai}},\ }\bibfield  {title} {\bibinfo {title} {Spin
  fluctuations in frustrated kagom\'e lattice system
  \text{Sr}\text{Cr$_{8}$}\text{Ga$_{4}$}\text{O$_{19}$} studied by muon spin
  relaxation},\ }\href {https://doi.org/10.1103/PhysRevLett.73.3306} {\bibfield
   {journal} {\bibinfo  {journal} {Phys. Rev. Lett.}\ }\textbf {\bibinfo
  {volume} {73}},\ \bibinfo {pages} {3306} (\bibinfo {year}
  {1994})}\BibitemShut {NoStop}%
\bibitem [{\citenamefont {Kanazawa}\ \emph {et~al.}(2011)\citenamefont
  {Kanazawa}, \citenamefont {Onose}, \citenamefont {Arima}, \citenamefont
  {Okuyama}, \citenamefont {Ohoyama}, \citenamefont {Wakimoto}, \citenamefont
  {Kakurai}, \citenamefont {Ishiwata},\ and\ \citenamefont
  {Tokura}}]{PhysRevLett.106.156603}%
  \BibitemOpen
  \bibfield  {author} {\bibinfo {author} {\bibfnamefont {N.}~\bibnamefont
  {Kanazawa}}, \bibinfo {author} {\bibfnamefont {Y.}~\bibnamefont {Onose}},
  \bibinfo {author} {\bibfnamefont {T.}~\bibnamefont {Arima}}, \bibinfo
  {author} {\bibfnamefont {D.}~\bibnamefont {Okuyama}}, \bibinfo {author}
  {\bibfnamefont {K.}~\bibnamefont {Ohoyama}}, \bibinfo {author} {\bibfnamefont
  {S.}~\bibnamefont {Wakimoto}}, \bibinfo {author} {\bibfnamefont
  {K.}~\bibnamefont {Kakurai}}, \bibinfo {author} {\bibfnamefont
  {S.}~\bibnamefont {Ishiwata}},\ and\ \bibinfo {author} {\bibfnamefont
  {Y.}~\bibnamefont {Tokura}},\ }\bibfield  {title} {\bibinfo {title} {Large
  topological hall effect in a short-period helimagnet mnge},\ }\href
  {https://doi.org/10.1103/PhysRevLett.106.156603} {\bibfield  {journal}
  {\bibinfo  {journal} {Phys. Rev. Lett.}\ }\textbf {\bibinfo {volume} {106}},\
  \bibinfo {pages} {156603} (\bibinfo {year} {2011})}\BibitemShut {NoStop}%
\bibitem [{\citenamefont {Battle}\ \emph {et~al.}(1986)\citenamefont {Battle},
  \citenamefont {Cheetham}, \citenamefont {Harrison},\ and\ \citenamefont
  {Long}}]{BATTLE198616}%
  \BibitemOpen
  \bibfield  {author} {\bibinfo {author} {\bibfnamefont {P.~D.}\ \bibnamefont
  {Battle}}, \bibinfo {author} {\bibfnamefont {A.~K.}\ \bibnamefont
  {Cheetham}}, \bibinfo {author} {\bibfnamefont {W.~T.}\ \bibnamefont
  {Harrison}},\ and\ \bibinfo {author} {\bibfnamefont {G.~J.}\ \bibnamefont
  {Long}},\ }\bibfield  {title} {\bibinfo {title} {The crystal structure and
  magnetic properties of the synthetic langbeinite
  \text{K}\text{Ba}\text{Fe$_{2}$}\text{(PO$_{4})$}$_{3}$},\ }\href
  {https://doi.org/https://doi.org/10.1016/0022-4596(86)90211-2} {\bibfield
  {journal} {\bibinfo  {journal} {J. Solid State Chem.}\ }\textbf {\bibinfo
  {volume} {62}},\ \bibinfo {pages} {16} (\bibinfo {year} {1986})}\BibitemShut
  {NoStop}%
\bibitem [{\citenamefont {Do}\ \emph {et~al.}(2018)\citenamefont {Do},
  \citenamefont {Lee}, \citenamefont {Lee}, \citenamefont {Choi}, \citenamefont
  {Lee}, \citenamefont {Gorbunov}, \citenamefont {Wosnitza}, \citenamefont
  {Suh},\ and\ \citenamefont {Choi}}]{PhysRevB.98.014407}%
  \BibitemOpen
  \bibfield  {author} {\bibinfo {author} {\bibfnamefont {S.-H.}\ \bibnamefont
  {Do}}, \bibinfo {author} {\bibfnamefont {W.-J.}\ \bibnamefont {Lee}},
  \bibinfo {author} {\bibfnamefont {S.}~\bibnamefont {Lee}}, \bibinfo {author}
  {\bibfnamefont {Y.~S.}\ \bibnamefont {Choi}}, \bibinfo {author}
  {\bibfnamefont {K.-J.}\ \bibnamefont {Lee}}, \bibinfo {author} {\bibfnamefont
  {D.~I.}\ \bibnamefont {Gorbunov}}, \bibinfo {author} {\bibfnamefont
  {J.}~\bibnamefont {Wosnitza}}, \bibinfo {author} {\bibfnamefont {B.~J.}\
  \bibnamefont {Suh}},\ and\ \bibinfo {author} {\bibfnamefont {K.-Y.}\
  \bibnamefont {Choi}},\ }\bibfield  {title} {\bibinfo {title} {Short-range
  quasistatic order and critical spin correlations in
  $\alpha$-\text{Ru}$_{1-x}$\text{Ir}$_x$\text{Cl}$_3$},\ }\href
  {https://doi.org/10.1103/PhysRevB.98.014407} {\bibfield  {journal} {\bibinfo
  {journal} {Phys. Rev. B}\ }\textbf {\bibinfo {volume} {98}},\ \bibinfo
  {pages} {014407} (\bibinfo {year} {2018})}\BibitemShut {NoStop}%
\bibitem [{\citenamefont {Peng}\ and\ \citenamefont
  {Zhang}(2025)}]{PhysRevB.111.014409}%
  \BibitemOpen
  \bibfield  {author} {\bibinfo {author} {\bibfnamefont {C.}~\bibnamefont
  {Peng}}\ and\ \bibinfo {author} {\bibfnamefont {L.}~\bibnamefont {Zhang}},\
  }\bibfield  {title} {\bibinfo {title} {Scaling and data collapse of
  two-dimensional random singlet states in a magnetic field},\ }\href
  {https://doi.org/10.1103/PhysRevB.111.014409} {\bibfield  {journal} {\bibinfo
   {journal} {Phys. Rev. B}\ }\textbf {\bibinfo {volume} {111}},\ \bibinfo
  {pages} {014409} (\bibinfo {year} {2025})}\BibitemShut {NoStop}%
\bibitem [{\citenamefont {Hong}\ \emph {et~al.}(2021)\citenamefont {Hong},
  \citenamefont {Liu}, \citenamefont {Liu}, \citenamefont {Ma}, \citenamefont
  {Koda}, \citenamefont {Li}, \citenamefont {Song}, \citenamefont {Yang},
  \citenamefont {Yang}, \citenamefont {Cheng}, \citenamefont {Zhang},
  \citenamefont {Bao}, \citenamefont {Ma}, \citenamefont {Chen}, \citenamefont
  {Sun}, \citenamefont {Guo}, \citenamefont {Luo}, \citenamefont {Sandvik},\
  and\ \citenamefont {Li}}]{PhysRevLett.126.037201}%
  \BibitemOpen
  \bibfield  {author} {\bibinfo {author} {\bibfnamefont {W.}~\bibnamefont
  {Hong}}, \bibinfo {author} {\bibfnamefont {L.}~\bibnamefont {Liu}}, \bibinfo
  {author} {\bibfnamefont {C.}~\bibnamefont {Liu}}, \bibinfo {author}
  {\bibfnamefont {X.}~\bibnamefont {Ma}}, \bibinfo {author} {\bibfnamefont
  {A.}~\bibnamefont {Koda}}, \bibinfo {author} {\bibfnamefont {X.}~\bibnamefont
  {Li}}, \bibinfo {author} {\bibfnamefont {J.}~\bibnamefont {Song}}, \bibinfo
  {author} {\bibfnamefont {W.}~\bibnamefont {Yang}}, \bibinfo {author}
  {\bibfnamefont {J.}~\bibnamefont {Yang}}, \bibinfo {author} {\bibfnamefont
  {P.}~\bibnamefont {Cheng}}, \bibinfo {author} {\bibfnamefont
  {H.}~\bibnamefont {Zhang}}, \bibinfo {author} {\bibfnamefont
  {W.}~\bibnamefont {Bao}}, \bibinfo {author} {\bibfnamefont {X.}~\bibnamefont
  {Ma}}, \bibinfo {author} {\bibfnamefont {D.}~\bibnamefont {Chen}}, \bibinfo
  {author} {\bibfnamefont {K.}~\bibnamefont {Sun}}, \bibinfo {author}
  {\bibfnamefont {W.}~\bibnamefont {Guo}}, \bibinfo {author} {\bibfnamefont
  {H.}~\bibnamefont {Luo}}, \bibinfo {author} {\bibfnamefont {A.~W.}\
  \bibnamefont {Sandvik}},\ and\ \bibinfo {author} {\bibfnamefont
  {S.}~\bibnamefont {Li}},\ }\bibfield  {title} {\bibinfo {title} {Extreme
  suppression of antiferromagnetic order and critical scaling in a
  two-dimensional random quantum magnet},\ }\href
  {https://doi.org/10.1103/PhysRevLett.126.037201} {\bibfield  {journal}
  {\bibinfo  {journal} {Phys. Rev. Lett.}\ }\textbf {\bibinfo {volume} {126}},\
  \bibinfo {pages} {037201} (\bibinfo {year} {2021})}\BibitemShut {NoStop}%
\bibitem [{\citenamefont {Shimokawa}\ \emph {et~al.}(2025)\citenamefont
  {Shimokawa}, \citenamefont {Sabharwal},\ and\ \citenamefont
  {Shannon}}]{shimokawah}%
  \BibitemOpen
  \bibfield  {author} {\bibinfo {author} {\bibfnamefont {T.}~\bibnamefont
  {Shimokawa}}, \bibinfo {author} {\bibfnamefont {S.}~\bibnamefont
  {Sabharwal}},\ and\ \bibinfo {author} {\bibfnamefont {N.}~\bibnamefont
  {Shannon}},\ }\href {https://arxiv.org/abs/2505.11874} {\bibinfo {title} {Can
  experimentally-accessible measures of entanglement distinguish quantum spin
  liquids from disorder-driven "random singlet" phases ?}} (\bibinfo {year}
  {2025}),\ \Eprint {https://arxiv.org/abs/2505.11874} {arXiv:2505.11874
  [cond-mat.str-el]} \BibitemShut {NoStop}%
\bibitem [{\citenamefont {Lee}\ \emph {et~al.}(2023)\citenamefont {Lee},
  \citenamefont {Lee}, \citenamefont {Kim}, \citenamefont {Kittaka},
  \citenamefont {Kohama}, \citenamefont {Sakakibara}, \citenamefont {Lee},
  \citenamefont {van Tol}, \citenamefont {Gorbunov}, \citenamefont {Do},
  \citenamefont {Yoon}, \citenamefont {Berlie},\ and\ \citenamefont
  {Choi}}]{PhysRevB.107.214404}%
  \BibitemOpen
  \bibfield  {author} {\bibinfo {author} {\bibfnamefont {C.}~\bibnamefont
  {Lee}}, \bibinfo {author} {\bibfnamefont {S.}~\bibnamefont {Lee}}, \bibinfo
  {author} {\bibfnamefont {H.-S.}\ \bibnamefont {Kim}}, \bibinfo {author}
  {\bibfnamefont {S.}~\bibnamefont {Kittaka}}, \bibinfo {author} {\bibfnamefont
  {Y.}~\bibnamefont {Kohama}}, \bibinfo {author} {\bibfnamefont
  {T.}~\bibnamefont {Sakakibara}}, \bibinfo {author} {\bibfnamefont {K.~H.}\
  \bibnamefont {Lee}}, \bibinfo {author} {\bibfnamefont {J.}~\bibnamefont {van
  Tol}}, \bibinfo {author} {\bibfnamefont {D.~I.}\ \bibnamefont {Gorbunov}},
  \bibinfo {author} {\bibfnamefont {S.-H.}\ \bibnamefont {Do}}, \bibinfo
  {author} {\bibfnamefont {S.}~\bibnamefont {Yoon}}, \bibinfo {author}
  {\bibfnamefont {A.}~\bibnamefont {Berlie}},\ and\ \bibinfo {author}
  {\bibfnamefont {K.-Y.}\ \bibnamefont {Choi}},\ }\bibfield  {title} {\bibinfo
  {title} {Random singlets in the $s=5/2$ coupled frustrated cubic lattice
  \text{Lu$_{3}$}\text{Sb$_{3}$}\text{Mn$_{2}$}\text{O$_{14}$}},\ }\href
  {https://doi.org/10.1103/PhysRevB.107.214404} {\bibfield  {journal} {\bibinfo
   {journal} {Phys. Rev. B}\ }\textbf {\bibinfo {volume} {107}},\ \bibinfo
  {pages} {214404} (\bibinfo {year} {2023})}\BibitemShut {NoStop}%
\bibitem [{\citenamefont {Silverman}\ and\ \citenamefont
  {Adler}(1990)}]{PhysRevB.42.1369}%
  \BibitemOpen
  \bibfield  {author} {\bibinfo {author} {\bibfnamefont {A.}~\bibnamefont
  {Silverman}}\ and\ \bibinfo {author} {\bibfnamefont {J.}~\bibnamefont
  {Adler}},\ }\bibfield  {title} {\bibinfo {title} {Site-percolation threshold
  for a diamond lattice with diatomic substitution},\ }\href
  {https://doi.org/10.1103/PhysRevB.42.1369} {\bibfield  {journal} {\bibinfo
  {journal} {Phys. Rev. B}\ }\textbf {\bibinfo {volume} {42}},\ \bibinfo
  {pages} {1369} (\bibinfo {year} {1990})}\BibitemShut {NoStop}%
\bibitem [{\citenamefont {Moessner}\ and\ \citenamefont
  {Chalker}(1998)}]{PhysRevB.58.12049}%
  \BibitemOpen
  \bibfield  {author} {\bibinfo {author} {\bibfnamefont {R.}~\bibnamefont
  {Moessner}}\ and\ \bibinfo {author} {\bibfnamefont {J.~T.}\ \bibnamefont
  {Chalker}},\ }\bibfield  {title} {\bibinfo {title} {Low-temperature
  properties of classical geometrically frustrated antiferromagnets},\ }\href
  {https://doi.org/10.1103/PhysRevB.58.12049} {\bibfield  {journal} {\bibinfo
  {journal} {Phys. Rev. B}\ }\textbf {\bibinfo {volume} {58}},\ \bibinfo
  {pages} {12049} (\bibinfo {year} {1998})}\BibitemShut {NoStop}%
\bibitem [{\citenamefont {Kermarrec}\ \emph {et~al.}(2021)\citenamefont
  {Kermarrec}, \citenamefont {Kumar}, \citenamefont {Bernard}, \citenamefont
  {H\'enaff}, \citenamefont {Mendels}, \citenamefont {Bert}, \citenamefont
  {Paulose}, \citenamefont {Hazra},\ and\ \citenamefont
  {Koteswararao}}]{PhysRevLett.127.157202}%
  \BibitemOpen
  \bibfield  {author} {\bibinfo {author} {\bibfnamefont {E.}~\bibnamefont
  {Kermarrec}}, \bibinfo {author} {\bibfnamefont {R.}~\bibnamefont {Kumar}},
  \bibinfo {author} {\bibfnamefont {G.}~\bibnamefont {Bernard}}, \bibinfo
  {author} {\bibfnamefont {R.}~\bibnamefont {H\'enaff}}, \bibinfo {author}
  {\bibfnamefont {P.}~\bibnamefont {Mendels}}, \bibinfo {author} {\bibfnamefont
  {F.}~\bibnamefont {Bert}}, \bibinfo {author} {\bibfnamefont {P.~L.}\
  \bibnamefont {Paulose}}, \bibinfo {author} {\bibfnamefont {B.~K.}\
  \bibnamefont {Hazra}},\ and\ \bibinfo {author} {\bibfnamefont
  {B.}~\bibnamefont {Koteswararao}},\ }\bibfield  {title} {\bibinfo {title}
  {Classical spin liquid state in the $s=\frac{5}{2}$ \text{H}eisenberg kagome
  antiferromagnet
  \text{Li$_{9}$}\text{Fe$_{3}$}(\text{P$_{2}$}\text{O$_{7}$})$_{3}$(\text{P}\text{O$_{4}$})$_{2}$},\
  }\href {https://doi.org/10.1103/PhysRevLett.127.157202} {\bibfield  {journal}
  {\bibinfo  {journal} {Phys. Rev. Lett.}\ }\textbf {\bibinfo {volume} {127}},\
  \bibinfo {pages} {157202} (\bibinfo {year} {2021})}\BibitemShut {NoStop}%
\bibitem [{\citenamefont {J.~Khatua}\ and\ \citenamefont
  {Krieger}()}]{psidatabase}%
  \BibitemOpen
  \bibfield  {author} {\bibinfo {author} {\bibfnamefont {K.-Y.~C.}\
  \bibnamefont {J.~Khatua}}\ and\ \bibinfo {author} {\bibfnamefont {J.~A.}\
  \bibnamefont {Krieger}},\ }\href@noop {} {\bibinfo {title} {\text{PSI}
  $\mu$\text{SR} experiment database}},\ \bibinfo {howpublished}
  {\url{http://musruser.psi.ch/cgi-bin/SearchDB.cgi}},\ \bibinfo {note}
  {accessed: [28/06/2025]}\BibitemShut {NoStop}%
\end{thebibliography}%
\end{document}